\documentclass[runningheads]{llncs}
\usepackage[T1]{fontenc}
\usepackage{hyperref}
\usepackage{url}
\usepackage{xcolor}
\usepackage{xspace}
\usepackage{makecell}
\usepackage{algorithm}
\usepackage{algpseudocode}
\usepackage{newfloat}
\usepackage{listings}
\usepackage{soul}
\usepackage{caption}
\usepackage{subfigure}
\usepackage{times}  
\usepackage{helvet}  
\usepackage{courier}  
\usepackage{bibentry}
\usepackage{graphicx}
\usepackage{multirow}
\usepackage{amsmath}
\usepackage{amsfonts,bm}
\usepackage{amssymb}
\begin{document}
\newcommand{\algo}[1]{\texttt{CDGraph}\xspace}
\title{\algo~: Dual Conditional Social Graph Synthesizing via Diffusion Model}
%
%
\author{Jui-Yi Tsai\inst{1}\and
Ya-Wen Teng\inst{2} \and
Ho Chiok-Yew\inst{3} \and
De-Nian Yang\inst{1} \and
Lydia Y. Chen\inst{4}}
\authorrunning{J.-Y. Tsai et al.}
%
\institute{Institute of Information Science, Academia Sinica, Taiwan \\
\email{\{vincenttsai,dnyang\}@iis.sinica.edu.tw} \and
Research Center for Information Technology Innovation, Academia Sinica, Taiwan \\
\email{ywteng@citi.sinica.edu.tw} \and
Department of Electrical Engineering, National Taiwan University, Taiwan \\
\email{cyho2212@gmail.com} \and
Department of Computer Science, TU Delft, The Netherlands \\
\email{lydiaychen@ieee.org}
}

\maketitle              
\begin{abstract}
The social graphs synthesized by the generative models are increasingly in demand due to data scarcity and concerns over user privacy. One of the key performance criteria for generating social networks is the fidelity to specified conditionals, such as users with certain membership and financial status. While recent diffusion models have shown remarkable performance in generating images, their effectiveness in synthesizing graphs has not yet been explored in the context of conditional social graphs. In this paper, we propose the first kind of conditional diffusion model for social networks, \algo~, which trains and synthesizes graphs based on two specified conditions. We propose the co-evolution dependency in the denoising process of \algo\ to capture the mutual dependencies between the dual conditions and further incorporate social homophily and social contagion to preserve the connectivity between nodes while satisfying the specified conditions. Moreover, we introduce a novel classifier loss, which guides the training of the diffusion process through the mutual dependency of dual conditions. We evaluate \algo\ against four existing graph generative methods, i.e., SPECTRE, GSM, EDGE, and DiGress, on four datasets. Our results show that the generated graphs from \algo\ achieve much higher dual-conditional validity and lower discrepancy in various social network metrics than the baselines, thus demonstrating its proficiency in generating dual-conditional social graphs.

\keywords{Social Network  \and Diffusion Model \and Graph Generation}
\end{abstract}
\section{Introduction}
Social networks offer a wide range of applications, such as viral marketing, friend recommendations, fake news detection, and more. However, achieving effective results often requires a substantial amount of personal data. Nevertheless, with the rise of privacy awareness, most individuals are reluctant to publicly disclose their personal information, including their profile and social interaction records, leading to a scarcity of data. The need of generating a social graph similar to the original one arises. It is critical for synthetic graphs to not only have similar structures as the original ones, e.g., centrality, but also satisfy exogenous conditions, such as {specific user profiles.} For example, when evaluating a customer's influence on a luxury golf club brand, it is more relevant to analyze a subset of her friends who have golf as a hobby, rather than analyzing her entire group of friends.  

Statistical sampling approaches~\cite{TBD2018Shuai,WWW2022Schweimer} have been used to produce graphs with certain social network properties, such as skewed degree distribution, a small diameter, and a large connected component, but they struggle to ensure the structure similarity to the original social graphs. Moreover, these methods cannot control the generation process to satisfy specified conditions (e.g., profiles of a social network user)~\cite{CSUR2020Bonifati}. Recently, deep generative models are shown effective for synthesizing molecular graphs \cite{JMLR2020Samanta,NIPS2020Chenthamarakshan}, via extracting latent features from input graphs. The deep molecular graphs~\cite{NIPS2022huang,ICLR2023vignac} can well preserve the network structure and deal with a single exogenous condition only (e.g., chemical properties such as toxicity, acidity, etc.) but not more. They fail to capture the dependency between two specified conditions, and thus cannot generate graphs satisfying both conditions. For instance, in the above example,~\cite{NIPS2022huang,ICLR2023vignac} may synthesize the graph satisfying the condition of users being golf enthusiasts, but they do not consider the income bracket relevant to purchasing power. 

\begin{figure}
    \centering
    \subfigure[Unconditional.]{%
        \centering
        \includegraphics[width=0.25\textwidth]{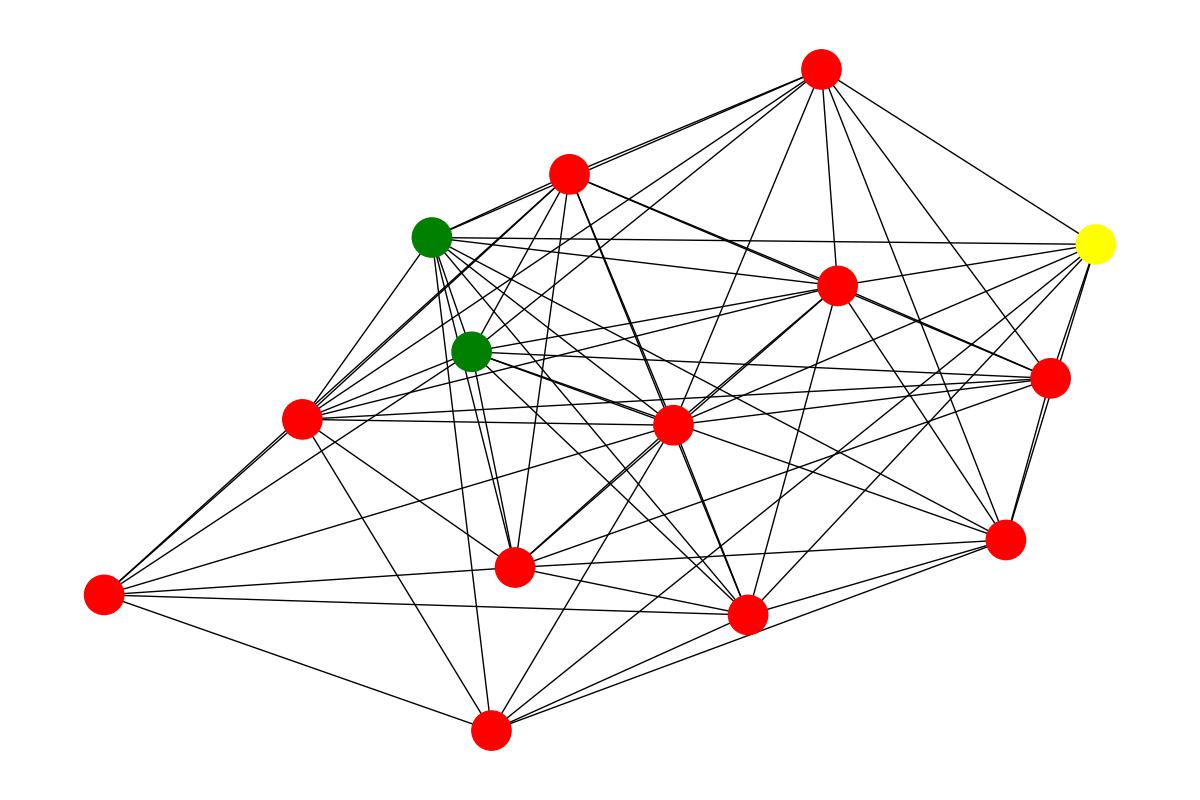}
        \label{fig:original}
    }\hfill
    \subfigure[ Single-conditional. ]{%
        \centering
        \includegraphics[width=0.25\textwidth]{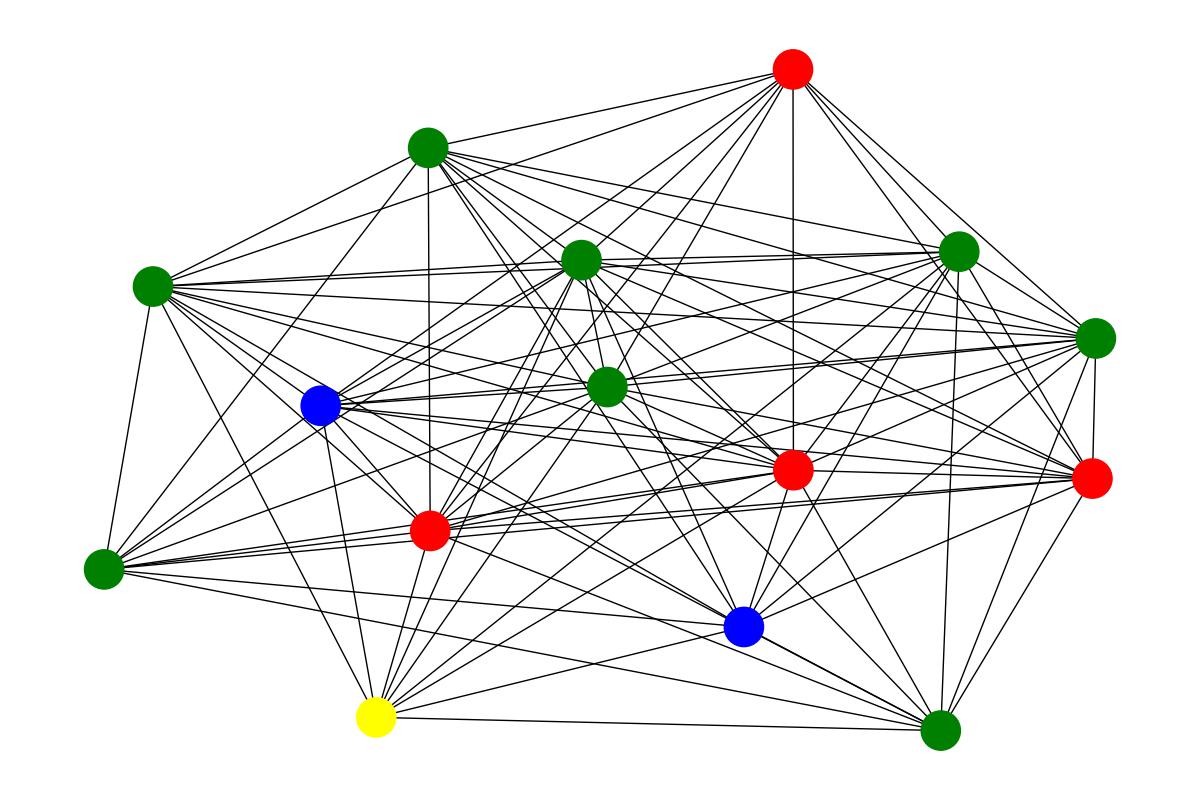}
        \label{fig:condition_golf}
    }\hfill
    \subfigure[Dual-conditional.]{%
        \centering
        \includegraphics[width=0.25\textwidth]{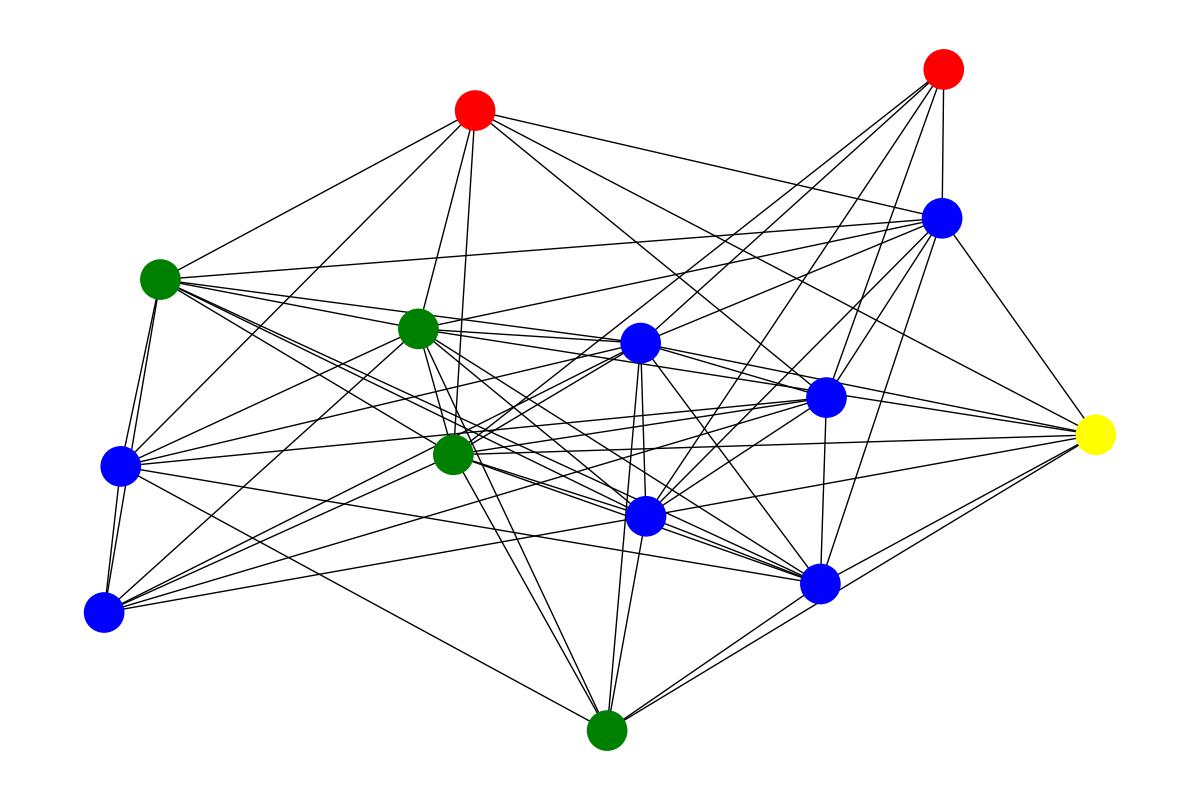}
        \label{fig:condition_both}
    }\hfill%
     \caption{Illustrative examples of synthetic social graphs generated by unconditional, single-conditional, and dual-conditional approaches. The colors of the node represent the condition satisfaction: none (red), one (green and yellow), and both (blue).}
     \label{fig:illustrative_example}
\end{figure}

Generating social graphs satisfying dual conditions is no mean feat and faces several challenges. 
(i)~\textit{Intricate dependencies across conditions}: Those exogenous conditions are often mutually dependent, and modeling conditions independently may lead to sub-optimal synthetic graphs. For instance, there is a high correlation between being golf enthusiasts and income brackets. (ii)~\textit{Fulfilling graph structure similarity and exogenous conditions}: When synthesizing social graphs, one has to not only maintain the network structure but also adhere to the conditions. Compared to chemical graphs, where the structure is dictated primarily by physical and chemical constraints, social homophily and social contagion are prevalent phenomena in social graphs. That is, users' profiles subtly drive their social interactions and vice versa. Hence, it is crucial to follow the original structure to generate and link users based on their profiles while ensuring that dual conditions are met (i.e., avoiding generating an excessive number of users who do not meet the specified conditions). Figure~\ref{fig:illustrative_example} illustrates the generated graphs obtained by unconditional, single-conditional, and dual-conditional generative models. It can be observed that both unconditional and single-conditional methods fall short in generating graphs that satisfy dual conditions, all the while maintaining a network structure influenced by social homophily and social contagion.

In this paper, we propose a novel conditional diffusion model, \textit{Dual Conditional Diffusion Graph (\algo~)}, for synthesizing social graphs based on two exogenous conditions jointly. We first propose a novel notion of {\textit{co-evolution dependency}} to capture the mutual dependencies between two exogenous conditions. In the conditional denoising process, we introduce the \textit{co-evolution dependency} to bind the diffusion processes to the specified node conditions. Moreover, the co-evolution dependency is designed to account for \textit{social homophily} and \textit{social contagion}, which explore the dependencies between the nodes' associated conditions and connections. The Social homophily-based co-evolution ensures nodes with similar profiles are connected, while the social contagion-based co-evolution facilitates nodes connected with edges to share similar profiles. Equipped with the co-evolution dependency, \algo\ can preserve the connectivity between nodes that satisfy the specified conditions.
Then, we design a novel loss function to train \algo\ with the guidance of the specified conditions, aiming to optimize the discrepancy in the co-evolution diffusion process. Furthermore, we introduce the notion of the \textit{dual-condition classifier} that jointly steers the co-evolution diffusion process toward the estimated distributions of condition fulfillment. 
We evaluate the performance of \algo~on real social networks by measuring the dual-conditional validity and the discrepancy in various social network metrics. The contributions include:

\begin{itemize}
    \item We derive the first kind of dual-conditional graph diffusion model, \algo~, for synthesizing social graphs. 
    
    \item We explore a novel feature of \textit{co-evolution dependency} incorporating \textit{social homophily} and \textit{social contagion} to capture the dependency between specified conditions so as to preserve the structural information between nodes satisfying specified conditions. 

    \item We introduce a novel loss function of the \textit{dual-condition classifier} that guides the denoising process of \algo~to jointly optimize the discrepancy in diffusion process and condition fulfillment.

    \item Through evaluations on four real-world social networks, we demonstrate that \algo\ outperforms the baselines in generating social graphs that satisfy the specified dual conditions and maintain social network properties.
\end{itemize}

\section{Related Work}
\textbf{Graph Generation.}
{The current research on graph generation techniques can be divided into statistic-based and deep generative model based ones. The existing statistic-based graph generation is mainly based on structural information such as network statistics{~\cite{WWW2022Schweimer}}, correlation~\cite{SIGMOD2015Erling}, community structure~\cite{TCSS2020Luo}, and node degree~\cite{ICDE2021Wang}, etc. There are several social graph generators for various purposes, e.g., frequent patterns~\cite{ICDM2013Shuai} and similarity across social network providers~\cite{TBD2018Shuai}.
{For deep generative model-based ones, they are based on auto-regression \cite{NIPS2019Liao,Shi2020ICLR}, variational autoencoder \cite{KDD2021Guo,JMLR2020Samanta}, and GAN \cite{pmlr2022martinkus}, etc. Especially, SPECTRE \cite{pmlr2022martinkus} is a GAN-based conditional generative model conditioning on graph Laplacian eigenvectors.}
However, both statistical and deep generative methods do not consider dual conditions, since they mainly simply focus on structural conditions or dependencies between consecutive steps, instead of the connectivity between users satisfying specified conditions (i.e., linking them according to the original structure), while avoiding to generate excessive irrelevant users for meaningful downstream analysis such as estimating social influence of a user to a specified population. 

\textbf{Diffusion Models.}
The current research direction on diffusion models mainly focuses on the application to multimedia, such as computer vision, text-image processing, and audio processing~\cite{NIPS2022Saharia,CVPR2022Gu,NIPS2021hoogeboom,ICLR2022Savinov}. 
Denoising Diffusion Probabilistic Model (DDPM)~\cite{NIPS2020Ho, NIPS2021Dhariwal} has performed significantly better than generative adversarial networks in image synthesis. Recently, diffusion models have been also applied to generate graph data (e.g., molecular graph generation) \cite{AISTATS2020niu,NIPS2022huang,ICLR2023vignac,chen2023ICML} thanks to the flexible modeling architecture and tractable probabilistic distribution compared with the aforementioned deep generative model architectures. In particular, DiGress \cite{ICLR2023vignac} synthesizes molecular graphs with the discrete denoising probabilistic model building on a discrete space. It exploits regression guidance to lead the denoising process to generate graphs that meet the condition property. 
EDGE \cite{chen2023ICML} is a discrete diffusion model exploiting graph sparsity to generate graphs conditioning on the change of node degree.  However, the above existing studies only deal with a single condition on nodes or edges in graphs, failing to capture the dependencies of conditions on graphs.

\section{\algo~}
In this section, {we begin by introducing the dual conditional graph generation problem. Subsequently, we revisit the concepts of the discrete diffusion model with a single condition of DiGress~\cite{ICLR2023vignac} as a preliminary to our approach.} 
Finally, we propose \algo~, featuring on the novel loss, the co-evolution dependency incorporating social contagion and social homophily, and the dual-condition classifier guidance.

We first provide the technical intuition behind \algo~. To guide the diffusion process with dual conditions, an intuitive approach is to extend the conditional DiGress~\cite{ICLR2023vignac} by adding the guidance of the second condition. However, the dependencies across conditions are not explicitly captured and fall short in guiding the graph generation. Hence, we introduce the aforementioned dependencies to jointly optimize the structural similarity and condition satisfaction.

\subsection{Formal Problem Definition}
Here, we formulate the problem of \textit{Dual Conditional Graph}. To perform the graph generation with the guidance of two conditions, we first define the \textit{ Condition Indication Graph} as follows. Figures~\ref{fig:condition_golf} and~\ref{fig:condition_both} illustrate the two examples of conditional indication graphs. 

\begin{definition}[Condition Indication Graph]
{Let $C$ denote the condition set. We define the \textit{condition indication graph} of $C$ by $G_{C}=(\{\mathbf{X}_{c}\}_{c \in C},\mathbf{E})$, where $\mathbf{X}_{c}$ indicate nodes' satisfaction of condition $c$, and $\mathbf{E}_{c}$ indicates the existence of edges between nodes. Specifically, $\mathbf{x}_{n,c} \in \{0,1\}^{2}$ in $\mathbf{X}_{c}$ is a one-hot encoding vector representing whether a node $v_{n}$ in $G_{C}$ satisfies condition $c$; $e_{n,m}=(v_{n},v_{m})$ in $\mathbf{E}$ is a one-hot encoding vector representing whether an edge between $v_{n}$ and $v_{m}$ in $G_{C}$ satisfies condition $c$.}
\end{definition}

\begin{definition}[Dual Conditional Graph Generation]
{Given the condition set $C=\{c_1,c_2\}$} and the condition indication graph $G_{C}=(\{\mathbf{X}_{c_{1}},\mathbf{X}_{c_{2}}\},\mathbf{E})$, the problem is to generate social graphs, such that i)~the structural information of the generated graphs is similar to $G_{C}$, and ii)~the majority of nodes in the generated graphs meet the conditions $c_{1}$ and $c_{2}$.
\end{definition}

\subsection{Discrete Diffusion Models for Graph Generation}
\subsubsection{Preliminary: Discretizing Diffusion Profess}
We revisit DiGress~\cite{ICLR2023vignac}, which is a discrete diffusion model with a single condition, i.e., $C=\{c\}$ and $G_C = (\{\mathbf{X}_{c}\},\mathbf{E})$. Typically, DiGress consists of two components: forward noising process and reverse denoising process.
For $t \geq 1$, the forward noising process of DiGress is defined by $q(G^{(t)}|G^{(t-1)})$ and $q(G^{(T)}|G^{(0)})=\prod_{t=1}^{T}q(G^{(t)}|G^{(t-1)})$.

For the reverse denoising process, given $G^{(t)}$, DiGress predicts the clean graph $G^{(0)}$ by a denoising neural network $\phi_{\theta}$ (parameterized by $\theta$) and obtains the reverse denoising process $p_{\theta}$ as follows:
\begin{align}
   \notag p_{\theta}(G^{(t-1)}|G^{(t)})&=q(G^{(t-1)}|G^{(t)},G^{(0)})p_{\theta}(G^{(0)}|G^{(t)});\\
    q(G^{(t-1)}|G^{(t)})&\propto q(G^{(t)}|G^{(t-1)})q(G^{(t-1)}|G^{(0)}),
\end{align}
in which $q(G^{(t-1)}|G^{(t)})$ can be approximated by the noising process.

To enable single conditional graph generation, DiGress guides the reverse denoising process by a machine learning model $f$ (i.e., a regression model) to push the predicted distribution toward graphs fulfilling the condition $c$. $f$ is trained to predict the condition $c$ of the input graph $G$ from the noised version $G^{(t)}$ such that $c\approx\hat{c}=f(G^{(t)})$. The reverse denoising process guided by a single condition is presented as follows:
\begin{align}
    q(G^{(t-1)}|G^{(t)},c)\propto q(c|G^{(t-1)})q(G^{(t-1)}|G^{(t)}),
\end{align}
where the first term is approximated by the learned distribution of the regression model, and the second term is approximated by the unconditional diffusion model.

\begin{figure}
    \centering
    \subfigure[Social Homophily]{%
        \centering
        \includegraphics[width=0.5\textwidth]{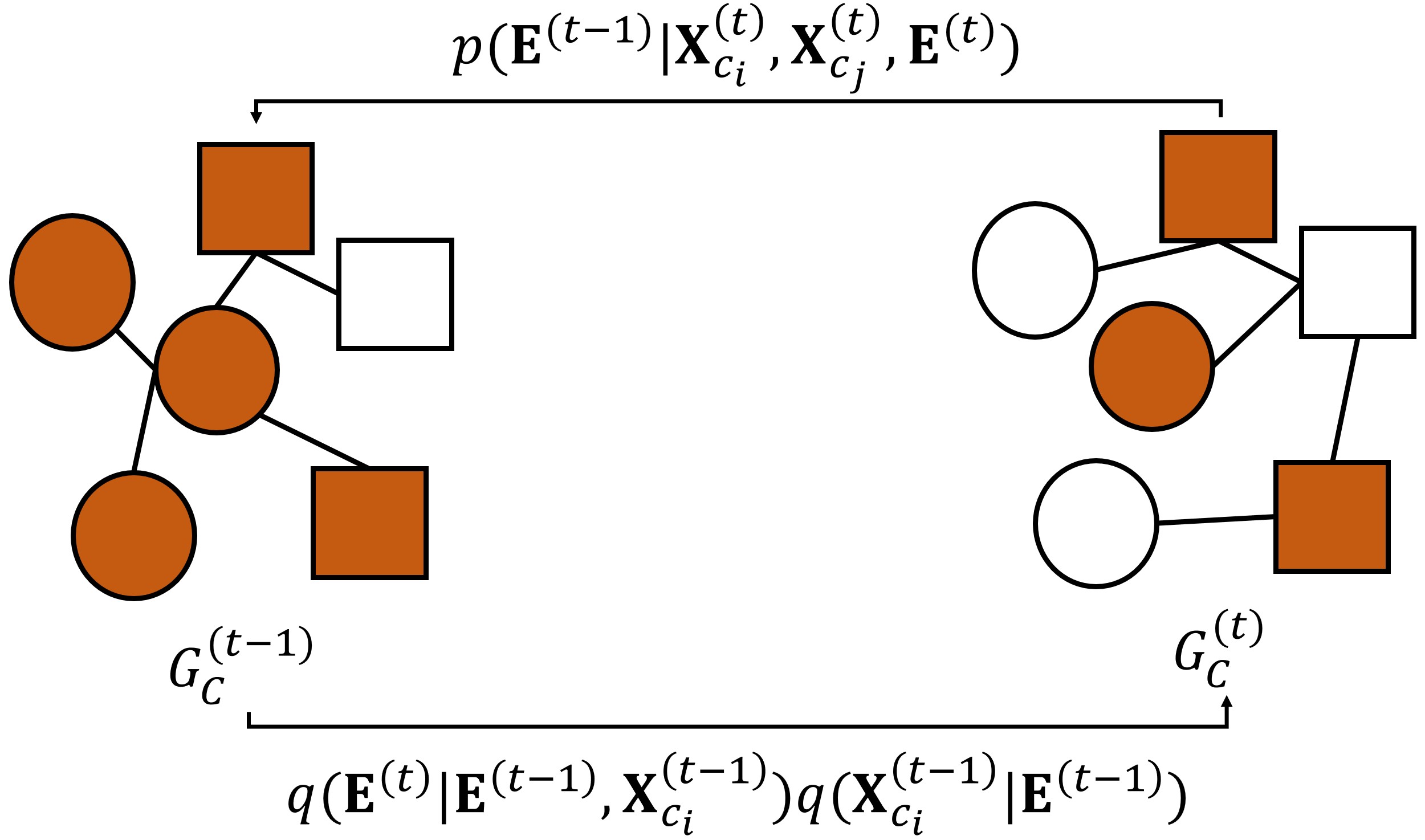}
        \label{fig:homophily}
    }%
    \subfigure[Social Contagion]{%
        \centering
        \includegraphics[width=0.5\textwidth]{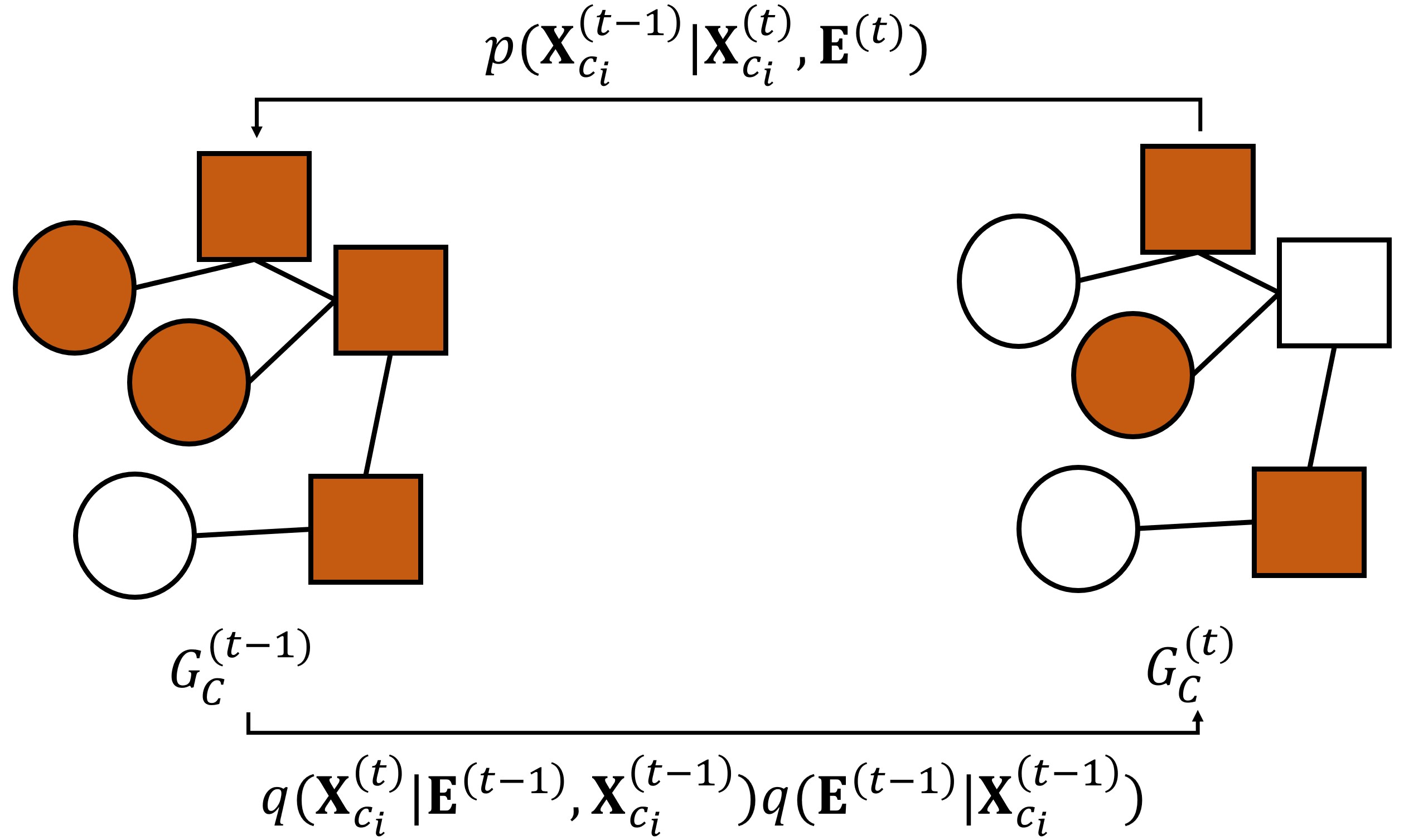}
        \label{fig:contagion}
    }
    \caption{Diffusion process in \algo~with notions of social homophily and social contagion.}
    \label{fig:notions}
\end{figure}

\subsection{Dual Conditional Graph Synthesizing}
Here, we present the overall learning framework of \algo~ and its novel features for integrating dual conditions. Specifically, to capture the dependency between two exogenous conditions, \algo~ introduces \textit{co-evolution dependency} to model the co-evolving diffusion process of dual conditions in the reverse denoising process with the notions of \textit{social homophily} and \textit{social contagion} so that the dependencies between node conditions and edge connections can be captured in the diffusion process. Then, to further fulfill dual conditions, we design a \textit{dual condition classifier} to guide the co-evolution diffusion process, modulating the estimated distribution to align with graphs satisfying specified conditions by the classifier loss. Specifically, \algo~comprises the forward noising process and the reverse denoising process. Given $C=\{c_1,c_2\}$ and the condition indication graph $G_C=(\{\mathbf{X}_{c_1},\mathbf{X}_{c_2}\},\mathbf{E})$, we model the forward noising processes for each of the specified conditions $c_{i}, \forall i \in \{1,2\}$ as follows: 
\begin{align}
    \notag&q(\mathbf{X}_{c_{i}}^{(t)}|\mathbf{X}_{c_{i}}^{(t-1)})=\mathbf{X}_{c_{i}}^{(t-1)}\mathbf{Q}_{X_{c_{i}}}^{(t)};q(\mathbf{E}^{(t)}|\mathbf{E}^{(t-1)})=\mathbf{E}^{(t-1)}\mathbf{Q}_{E}^{(t)},
\end{align}
where $\mathbf{Q}_{X_{c_{i}}}^{(t)}$ and $\mathbf{Q}_{E}^{(t)}$ are transition matrices for $\mathbf{X}_{c_{i}}$ and $\mathbf{E}$, respectively. Then we can show that $q(\mathbf{X}_{c_{i}}^{(t)}|\mathbf{X}_{c_{i}}^{(t-1)})$ and $q(\mathbf{E}^{(t)}|\mathbf{E}^{(t-1)})$ obey Bernoulli distributions as follows: 
\begin{align}
    q(\mathbf{X}_{c_{i}}^{(t)}|\mathbf{X}_{c_{i}}^{(t-1)})&=\mathcal{B}(\mathbf{X}_{c_{i}}^{(t)};(1-\beta_{t})\mathbf{X}_{c_{i}}^{(t-1)}+\beta_{t}\mathbf{1}/2);\\
    q(\mathbf{E}^{(t)}|\mathbf{E}^{(t-1)})&=\mathcal{B}(\mathbf{E}^{(t)};(1-\beta_{t})\mathbf{E}^{(t-1)}+\beta_{t}\mathbf{1}/2).
\end{align}
The detailed derivations regarding $\mathbf{Q}_{X_{c_{i}}}^{(t)}$ and $\mathbf{Q}_{E}^{(t)}$ are provided in Appendix~\ref{apdx:diffusion_process_derivation}.

The notions of \textit{social homophily} and \textit{social contagion} in the diffusion process in \algo~are illustrated in Figure~\ref{fig:notions}. In Fig.~\ref{fig:homophily}, as denoising from $G_{C}^{(t)}$ to $G_{C}^{(t-1)}$, the nodes with similar attributes (orange nodes) in $G_{C}^{(t-1)}$ tend to have links between them. In Fig.~\ref{fig:contagion}, as denoising from $G_{C}^{(t)}$ to $G_{C}^{(t-1)}$, the edges incident to orange-colored nodes tend to cause the other node have the same attribute.

\begin{figure}
    \centering
    \includegraphics[width=0.85\textwidth]{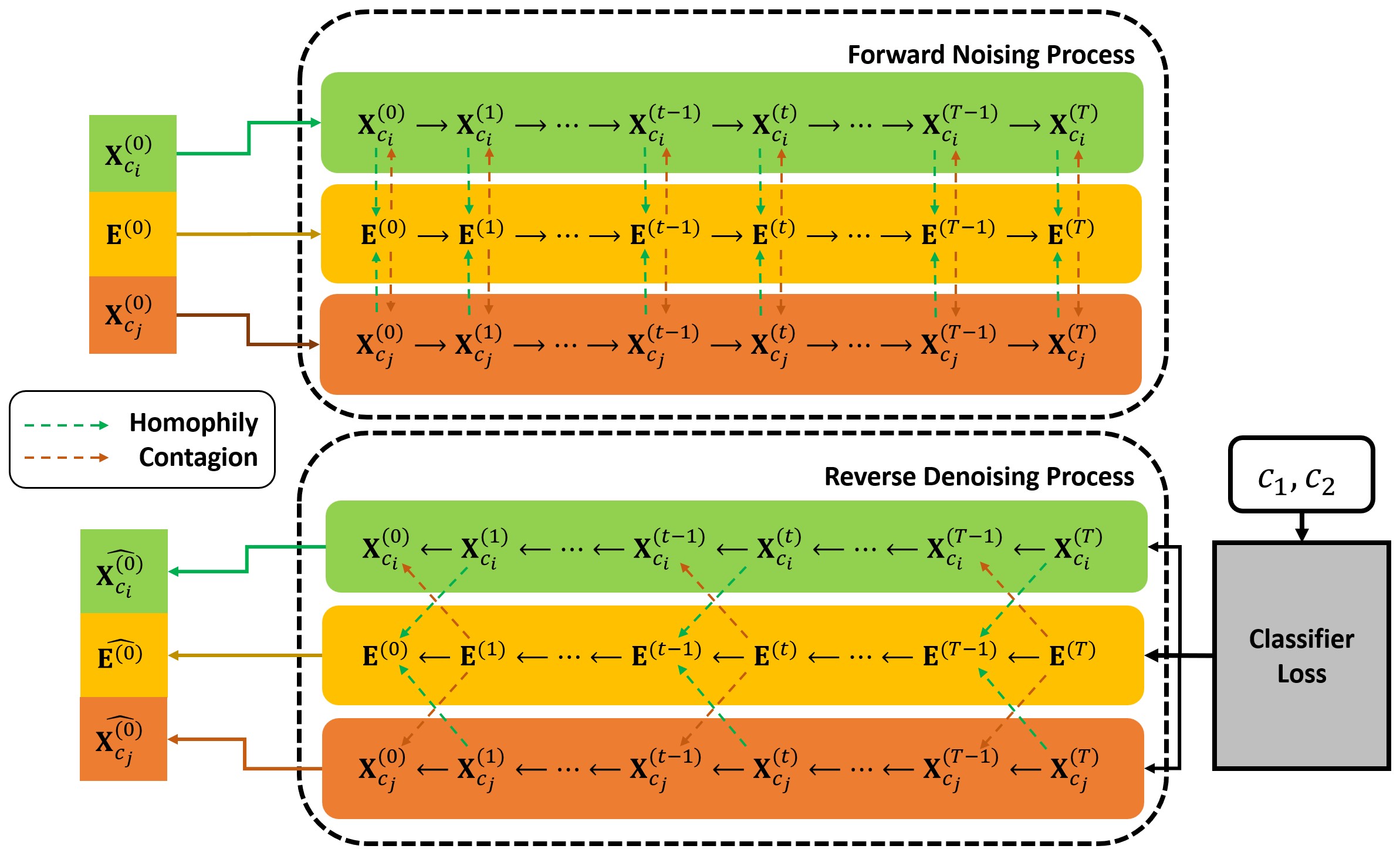}
    \caption{Workflow of \algo~.} 
    \label{fig:workflow2}
\end{figure}

By considering nodes with two conditions and the dependency between nodes and edges, the overall forward processes of $\mathbf{X}_{c_i}$ and $\mathbf{E}$ are formulated as follows:
\begin{align}
    \notag q(\mathbf{X}_{c_{i}}^{(0:T)})&=\prod_{t=1}^{T}q(\mathbf{X}_{c_{i}}^{(t)}|\mathbf{X}_{c_{i}}^{(t-1)},\mathbf{E}^{(t-1)})q(\mathbf{E}^{(t-1)}|\mathbf{X}_{c_{i}}^{(t-1)}),\\
    q(\mathbf{E}^{(0:T)})&=\prod_{t=1}^{T}q(\mathbf{E}^{(t)}|\mathbf{E}^{(t-1)},\mathbf{X}_{c_{i}}^{(t-1)})q(\mathbf{X}_{c_{i}}^{(t-1)}|\mathbf{E}^{(t-1)}).
\end{align}
The concept of the above process is illustrated in the upper half of Figure~\ref{fig:workflow2}, in which the green dashed lines represent the \textit{social homophily}, whereas the brown dashed lines represent the \textit{social contagion}. The former facilitates the connections between nodes with similar conditions; the latter makes adjacent nodes to have similar conditions. The detailed forward transition distribution with dependency in the above processes is defined and analyzed in Appendix~\ref{apdx:diffusion_process_derivation}.

\subsubsection{Co-evolution Dependency}
Different from DiGress, \algo~exploits \textit{co-evolution dependency} to model the intricate dependency across conditions of the nodes and connections between them in the denoising process, which is detailed in Appendix~\ref{apdx:diffusion_process_derivation}. This emphasis on capturing the relationship between $\mathbf{X}_{c_{i}}^{(t)}, i \in \{1,2\}$ and $\mathbf{E}^{(t)}$ are crucial for the precise reconstruction of the input graphs. To capture the dependencies between the connection between nodes and condition satisfaction of the nodes in a social graph, we build a denoising model incorporating two phenomena in social networks: \textit{social homophily} and \textit{social contagion}. The concept of the above denoising process is illustrated in the lower half of Figure~\ref{fig:workflow2}.

Assuming that $\mathbf{X}_{c_{i}}^{(t-1)}$, $\mathbf{X}_{c_{j}}^{(t-1)}$ and $\mathbf{E}^{(t-1)}$ are conditionally independent given $\mathbf{X}_{c_{i}}^{(t)}$, $\mathbf{X}_{c_{j}}^{(t)}$ and $\mathbf{E}^{(t)}$, the reverse denoising process can be further decomposed as follows:
\begin{align}
    \notag&p_{\theta}(\mathbf{X}_{c_{i}}^{(t-1)},\mathbf{X}_{c_{j}}^{(t-1)},\mathbf{E}^{(t-1)}|\mathbf{X}_{c_{i}}^{(t)},\mathbf{X}_{c_{j}}^{(t)},\mathbf{E}^{(t)})\\
    \notag&=p_{\theta}(\mathbf{X}_{c_{i}}^{(t-1)}|\mathbf{X}_{c_{i}}^{(t)},\mathbf{E}^{(t)})p_{\theta}(\mathbf{X}_{c_{j}}^{(t-1)}|\mathbf{X}_{c_{j}}^{(t)},\mathbf{E}^{(t)})p_{\theta}(\mathbf{E}^{(t-1)}|\mathbf{X}_{c_{i}}^{(t)},\mathbf{E}^{(t)},\mathbf{X}_{c_{j}}^{(t)}),\\
\end{align}
in which the first two terms represent \textit{social contagion} that can be used to denoise node conditions from given edges, and the third term represents \textit{social homophily} that can be exploited to denoise connections between nodes from given conditions of nodes. Note that $\mathbf{X}_{c_{i}}^{(t-1)}$ is independent on $\mathbf{X}_{c_{j}}^{(t)}$ (and $\mathbf{X}_{c_{j}}^{(t-1)}$ is independent on $\mathbf{X}_{c_{i}}^{(t)}$) for distinct $c_{i}$ and $c_{j}$.

\paragraph{Social Homophily-based Co-evolution.} 
In this paragraph, we discuss how to guide the diffusion process with social homophily, which states that nodes with similar conditions tend to have edges between them. Accordingly, we consider the following denoising process:
\begin{align}
    \notag&p_{\theta}(\mathbf{E}^{(0:T)})=p_{\theta}(\mathbf{X}_{c_{i}}^{(T)})\prod_{t=1}^{T}p_{\theta}(\mathbf{E}^{(t-1)}|\mathbf{E}^{(t)},\mathbf{X}_{c_{i}}^{(t)});\\
    &p_{\theta}(\mathbf{E}^{(t-1)}|\mathbf{E}^{(t)},\mathbf{X}_{c_{i}}^{(t)})=\mathcal{B}(\mathbf{p}_{\theta}^{(homo)}),
\end{align}
where
\begin{align*}
\mathbf{p}_{\theta}^{(homo)}=\sum_{\hat{\mathbf{E}}^{(0)} \in \{0,1\}}q(\mathbf{E}^{(t-1)}|\mathbf{E}^{(t)},\hat{\mathbf{E}}^{(0)})\hat{p}_{e}(\hat{\mathbf{E}}^{(0)}|\mathbf{E}^{(t)},\mathbf{X}_{c_{i}}^{(t)}),
\end{align*}
and $\hat{p}_{e}$ is the distribution learned to predict $\mathbf{E}^{(0)}$ from $\mathbf{E}^{(t)}$ conditioned on $\mathbf{X}_{c_{i}}^{(t)},\mathbf{X}_{c_{i}}^{(t)}$ by denoising network $\phi_{\theta}$.

The loss function of social homophily-based co-evolving diffusion can be derived as follows:
\begin{align}
    \notag&\mathcal{L}_{homo}=\sum_{t=2}^{T-1}D_{KL}[q(\mathbf{E}^{(t-1)}|\mathbf{E}^{(t)},\mathbf{E}^{(0)})\|p_{\theta}(\mathbf{E}^{(t-1)}|\mathbf{E}^{(t)},\mathbf{X}_{c_{i}}^{(t)},\mathbf{X}_{c_{j}}^{(t)})]\\
    &+D_{KL}[q(\mathbf{E}^{(T)}|\mathbf{E}^{(0)})\|p(\mathbf{E}^{(T)})]-\log p_{\theta}(\mathbf{E}^{(0)}|\mathbf{E}^{(1)},\mathbf{X}_{c_{i}}^{(1)},\mathbf{X}_{c_{j}}^{(1)}),
\end{align}
where the first term is the loss for diffusion process; the second term is the loss for prior distribution; the third term is the loss for reconstruction. 
\paragraph{Social Contagion-based Co-evolution.} 
In this paragraph, we discuss how to guide the diffusion process with social contagion, which states that nodes connected with edges tend to have similar conditions. Accordingly, we consider the following denoising process:
\begin{align}
    \notag&p_{\theta}(\mathbf{X}_{c_{i}}^{(0:T)})=p_{\theta}(\mathbf{E}^{(T)})\prod_{t=1}^{T}p_{\theta}(\mathbf{X}_{c_{i}}^{(t-1)}|\mathbf{X}_{c_{i}}^{(t)},\mathbf{E}^{(t)});\\
    &p_{\theta}(\mathbf{X}_{c_{i}}^{(t-1)}|\mathbf{X}_{c_{i}}^{(t)},\mathbf{E}^{(t)})=\mathcal{B}(\mathbf{p}_{\theta}^{(cont)}),
\end{align} 
where
\begin{align*}
\mathbf{p}_{\theta}^{(cont)}=\sum_{\hat{\mathbf{X}}_{c_{i}}}q(\mathbf{X}_{c_{i}}^{(t-1)}|\mathbf{X}_{c_{i}}^{(t)},\hat{\mathbf{X}}_{c_{i}}^{(0)})\hat{p}_{c_{i}}(\hat{\mathbf{X}}_{c_{i}}^{(0)}|\mathbf{X}_{c_{i}}^{(t)},\mathbf{E}^{(t)}),
\end{align*}
and $\hat{p}_{c_{i}}$ is the distribution learned to predict $\mathbf{X}_{c_{i}}^{(0)}$ from $\mathbf{X}_{c_{i}}^{(t)}$ conditioned on $\mathbf{E}^{(t)}$ by denoising network $\phi_{\theta}$.

And the loss function for social contagion-based co-evolution can be derived as follows:
\begin{align}
    \notag&\mathcal{L}_{cont}=\sum_{c_{i} \in C}\sum_{t=2}^{T-1}D_{KL}[q(\mathbf{X}_{c_{i}}^{(t-1)}|\mathbf{X}_{c_{i}}^{(t)},\mathbf{X}_{c_{i}}^{(0)})\|p_{\theta}(\mathbf{X}_{c_{i}}^{(t-1)}|\mathbf{X}_{c_{i}}^{(t)},\mathbf{E}^{(t)})]\\
    &+D_{KL}[q(\mathbf{X}_{c_{i}}^{(T)}|\mathbf{X}_{c_{i}}^{(0)})\|p(\mathbf{X}_{c_{i}}^{(T)})]-\log p_{\theta}(\mathbf{X}_{c_{i}}^{(0)}|\mathbf{X}_{c_{i}}^{(1)},\mathbf{E}^{(1)}).
\end{align}

The overall loss function of the co-evolution diffusion process is $\mathcal{L}=\mathcal{L}_{homo}+\mathcal{L}_{cont}$, which jointly optimizes the discrepancy in diffusion process and graph reconstruction in order to synthesize graphs with properties of social homophily and social contagion. The pseudocode of training and sampling is depicted in Appendix~\ref{apdx:pseudocode}.

\subsubsection{Dual-condition Classifier}
Afterward, \algo~leverages \textit{dual conditional classifier} to enable joint guidance for $p_{\theta}(G_{C}^{(t-1)}|G_{C}^{(t)})$ to fulfill dual conditions, instead of single conditional guidance in DiGress. To guide the diffusion process with two specified conditions jointly, we exploit the concept of conditional guidance and model the guidance distribution for the specified conditions $c_{i}$ and $c_{j}$ ($i \neq j$) with a classifier such that the generated graphs are classified according to whether a majority of the nodes satisfy both of the specified conditions or not. Note that the diffusion process satisfies the Markovian property:  $q({G}_{C}^{(t-1)}|{G}_{C}^{(t)},c_{i})=q({G}_{C}^{(t-1)}|{G}_{C}^{(t)}), \forall i$. 

Thus, from the results derived in the single-conditional denoising process, the co-evolution reverse denoising process with dual-condition classifier can be derived as follows:
\begin{align}
    &\notag q(G_{C}^{(t-1)}|G_{C}^{(t)},c_{i},c_{j})=\frac{q(c_{i}|G_{C}^{(t-1)},G_{C}^{(t)},c_{j})q(G_{C}^{(t-1)},G_{C}^{(t)},c_{j})}{q(c_{i}|G_{C}^{(t)},c_{j})q(c_{j}|G_{C}^{(t)})q(G_{C}^{(t)})}\\
    \notag&=\frac{q(c_{i}|G_{C}^{(t-1)},G_{C}^{(t)},c_{j})q(G_{C}^{(t-1)}|G_{C}^{(t)},c_{j})}{q(c_{i}|G_{C}^{(t)},c_{j})}\propto q(c_{i}|G_{C}^{(t-1)},c_{j})q(c_{j}|G_{C}^{(t-1)})q(G_{C}^{(t-1)}|G_{C}^{(t)}),
\end{align}
where the first two terms enable the hierarchical guidance of the conditions to guide the denoised graph $G_{C}^{(t-1)}$ satisfying one condition $c_{j}$ (i.e., more than half of the nodes in $G_{C}^{(t-1)}$ satisfy condition $c_{j}$), and then satisfying the other condition $c_{i}$ (i.e., more than half of the nodes in $G_{C}^{(t-1)}$ satisfy condition $c_{i}$) given that $G_{C}^{(t-1)}$ satisfies condition $c_{j}$. The third term $q(G_{C}^{(t-1)}|G_{C}^{(t)})$ is approximated by the aforementioned co-evolution denoising process with \textit{social homophily} and \textit{social contagion}. Note that, compared to DiGress and other previous studies, the proposed guidance model has more capability to guide the denoising process such that the denoised graphs meet both of the specified conditions if the specified conditions have a negative correlation. The pseudocode of conditional sampling is depicted in Appendix~\ref{apdx:pseudocode}.

\section{Empirical Evaluation}
\subsection{Setup}
\textbf{Datasets}. We conduct the experiments on the following datasets shown in Table~\ref{tab:datasets}. We sample ego networks with at maximum 100 nodes for each dataset (Facebook: 3548 graphs; Twitter: 75940 graphs; Flickr: 6450 graphs; BlogCatalog: 4219 graphs). 

\begin{table}[t]
    \centering
    \caption{Dataset descriptions}
    \scalebox{0.8}{
    \begin{tabular}{l c c c c}
        \hline
         & Facebook & Twitter & Flickr & BlogCatalog \\\hline
        \#Node & 4.03K & 81.3K & 7.57K & 5.19K \\\hline
        \#Edges & 88.23K & 216.83K & 479.47K & 343.48K \\\hline
    \end{tabular}
    }
    \label{tab:datasets}
\end{table}

\textbf{Baselines}. We compare \algo\ with the following baselines. (1) SPECTRE~\cite{pmlr2022martinkus}: A GAN-based graph generator modeling dominant components and generating graphs conditioning on graph Laplacian eigenvectors. (2) GSM~\cite{AISTATS2020niu}: A permutation invariant score-based graph generator modeling the gradient of the data distribution at the input graph with a multi-channel GNN architecture. (3) EDGE~\cite{chen2023ICML}: A discrete diffusion model that explicitly generates graphs conditioned on changes in node degree. (4) DiGress~\cite{ICLR2023vignac}: A discrete diffusion model that generates graphs by a categorical distribution of node/edge types (also with a regression guidance condition).\footnote{Since DiGress only considers a single condition, we extend it to handle the dual conditions by multiplying the guidance models of the respective conditions.}

\textbf{Evaluation Metrics}. We conduct comparative studies by evaluating the following metrics. (1) Validity (the higher, the better): It evaluates the proportion of generated graphs with a majority of nodes meeting both specified conditions. (2) Relative error ratios (the lower, the better): It evaluates the relative differences of the average numbers of nodes and edges, as well as the average density, between the input and generated graphs. (3) MMD (maximum mean discrepancy; the lower, the better): It evaluates the discrepancy in the distributions of clustering coefficients between the input and generated graphs.\footnote{Since we are primarily concerned with users who meet the specified conditions, we consider the subgraphs induced by those nodes that satisfy both specified conditions as the input graphs when evaluating the relative error ratios and MMD.}

\begin{table}[t]
    \centering
    \caption{Feasibility when the dual conditions are weakly positively correlated.}
    \scriptsize
    \begin{tabular}{l | c c c c c | c c c c c }
        \hline
        Dataset & \multicolumn{5}{c|}{Facebook} & \multicolumn{5}{c}{BlogCatalog} \\ \hline
        Metric & Validity & Node & Edge & Density & Clust. coeff. & Validity & Node & Edge & Density & Clust. coeff. \\\hline
        SPECTRE & \underline{0.448} & 0.243 & 1.795 & 1.119 & 0.313 & 0.3 & 0.026 & 6.081 & 5.494 & 0.098 \\\hline
        GSM & 0.272 & \textbf{0.100} & \underline{0.551} & \underline{0.392} & 0.031 & 0.29 & \textbf{0.013} & 3.486 & 3.441 & 0.063 \\\hline
        EDGE & 0.262 & 0.736 & 1.983 & 0.815 & \underline{0.028} & 0.292 & 0.414 & 0.510 & \underline{0.256} & \textbf{0.014} \\\hline
        DiGress & 0.375 & \underline{0.111} & 0.852 & 0.457 & 0.143 & \underline{0.3125} & 0.164 & \underline{0.413} & 0.699 & 0.266 \\\hline
        CDGraph & \textbf{1} & 0.149 & \textbf{0.186} & \textbf{0.002} & \textbf{0.022} & \textbf{1} & \underline{0.020} & \textbf{0.259} & \textbf{0.219} & \underline{0.031} \\\hline\hline
        Dataset & \multicolumn{5}{c|}{Twitter} & \multicolumn{5}{c}{Flickr} \\ \hline
        Metric & Validity & Node & Edge & Density & Clust. coeff. & Validity & Node & Edge & Density & Clust. coeff. \\\hline
        SPECTRE & 0.239 & 0.163 & 1.061 & \underline{0.131} & 0.359 & 0.323 & 0.053 & 2.115 & 2.275 & 1.006 \\\hline
        GSM & 0.309 & 0.205 & 0.417 & \textbf{0.034} & \underline{0.011} & 0.269 & \textbf{0.005} & 1.055 & 1.039 & \textbf{0.014} \\\hline
        EDGE & 0.291 & 0.101 & \underline{0.034} & 0.302 & 0.002 & 0.286 & 0.439 & 0.809 & 0.695 & \textbf{0.014} \\\hline
        DiGress & \underline{0.5} & \textbf{0.029} & 0.301 & 0.184 & 0.174 & \underline{0.375} & 0.090 & \underline{0.233} & \underline{0.139} & 0.286 \\\hline
        CDGraph & \textbf{1} & \underline{0.045} & \textbf{0.001} & 0.409 & \textbf{0.016} & \textbf{1} & \underline{0.008} & \textbf{0.056} & \textbf{0.068} & \underline{0.068} \\\hline
    \end{tabular}
    \label{tab:pos_summary}
\end{table}

\begin{table}[t]
    \centering
    \caption{Feasibility when the dual conditions are weakly negatively correlated.}
    \scriptsize
    \begin{tabular}{l | c c c c c | c c c c c }
        \hline
        Dataset & \multicolumn{5}{c|}{Facebook} & \multicolumn{5}{c}{BlogCatalog} \\ \hline
        Metric & Validity & Node & Edge & Density & Clust. coeff. & Validity & Node & Edge & Density & Clust. coeff. \\\hline
        SPECTRE & 0.241 & 0.241 & \underline{0.323} & 0.254 & 0.307 & 0.2 & 0.066 & 4.271 & 4.260 & 0.091 \\\hline
        GSM & 0.207 & \underline{0.176} & 0.837 & 0.506 & 0.032 & 0.244 & \underline{0.011} & 3.462 & 3.430 & 0.074 \\\hline
        EDGE & 0.256 & 0.805 & 2.135 & 0.852 & \underline{0.03} & 0.276 & 0.375 & \underline{0.513} & \underline{0.299} & \textbf{0.013} \\\hline
        DiGress & \underline{0.375} & 0.341 & 0.545 & \underline{0.093} & 0.361 & 0.438 & \textbf{0.005} & 1.066 & 1.028 & 0.263 \\\hline
        CDGraph & \textbf{0.94} & \textbf{0.102} & \textbf{0.261} & \textbf{0.060} & \textbf{0.016} & \textbf{1} & 0.020 & \textbf{0.259} & \textbf{0.219} & \underline{0.031} \\\hline
    \end{tabular}
    \label{tab:neg_summary}
\end{table}

\begin{figure}[t]
    \centering
    \subfigure[Input graph.]{%
        \centering
        \includegraphics[width=0.3\textwidth]{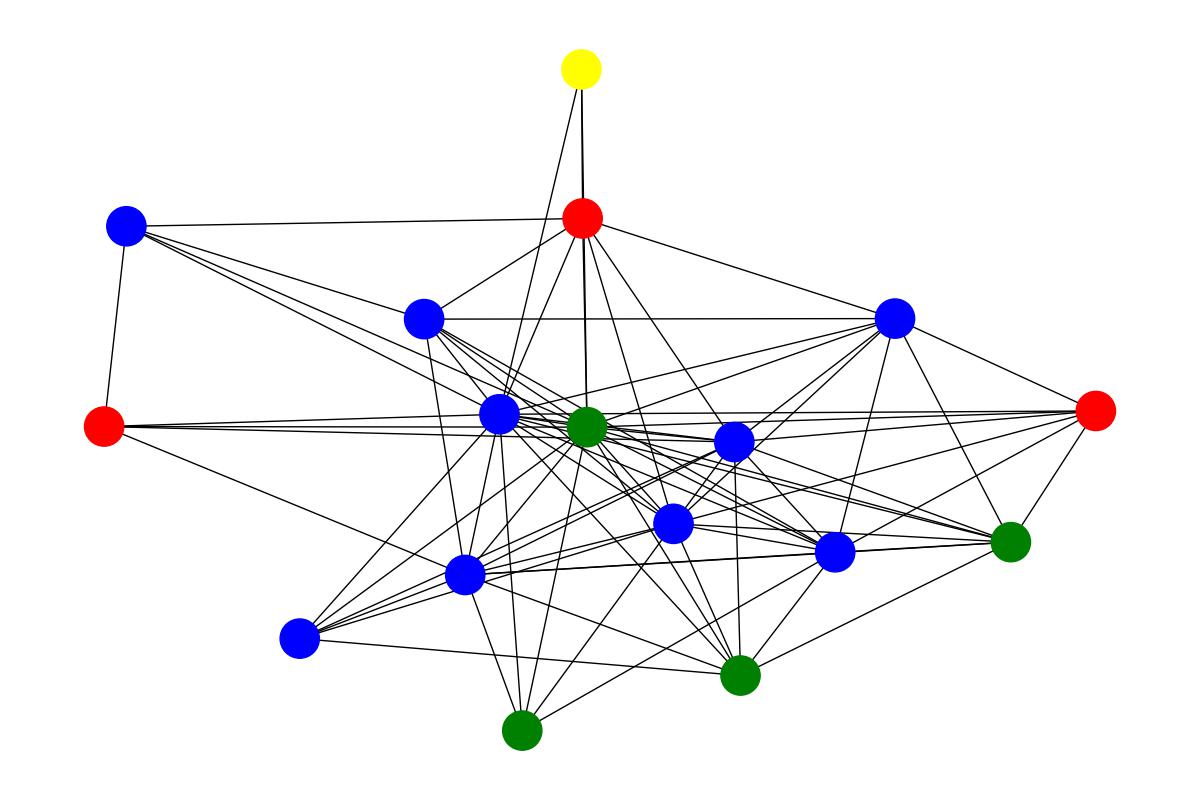}
        \label{fig:visual_tw_input}
    }\hfill
    \subfigure[Generation of SPECTRE.]{%
        \centering
        \includegraphics[width=0.3\textwidth]{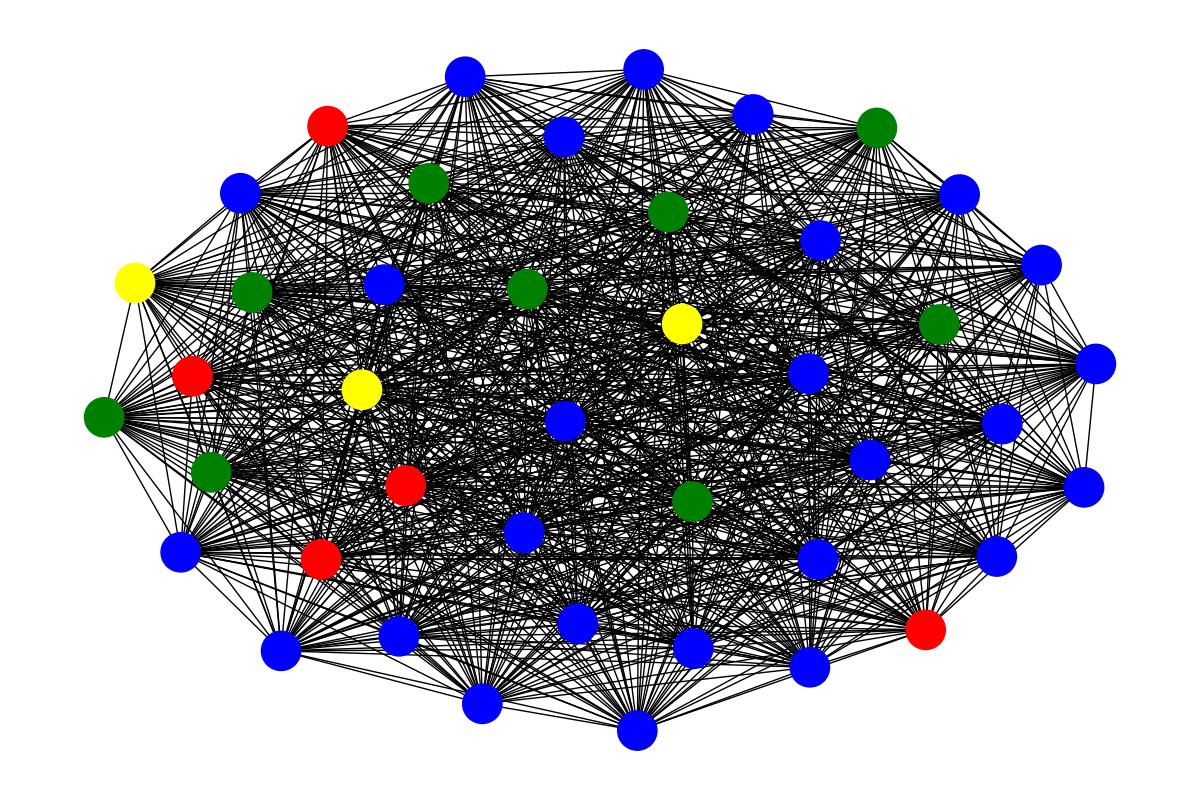}
        \label{fig:visual_tw_spectre}
    }\hfill
    \subfigure[Generation of GSM.]{%
        \centering
        \includegraphics[width=0.3\textwidth]{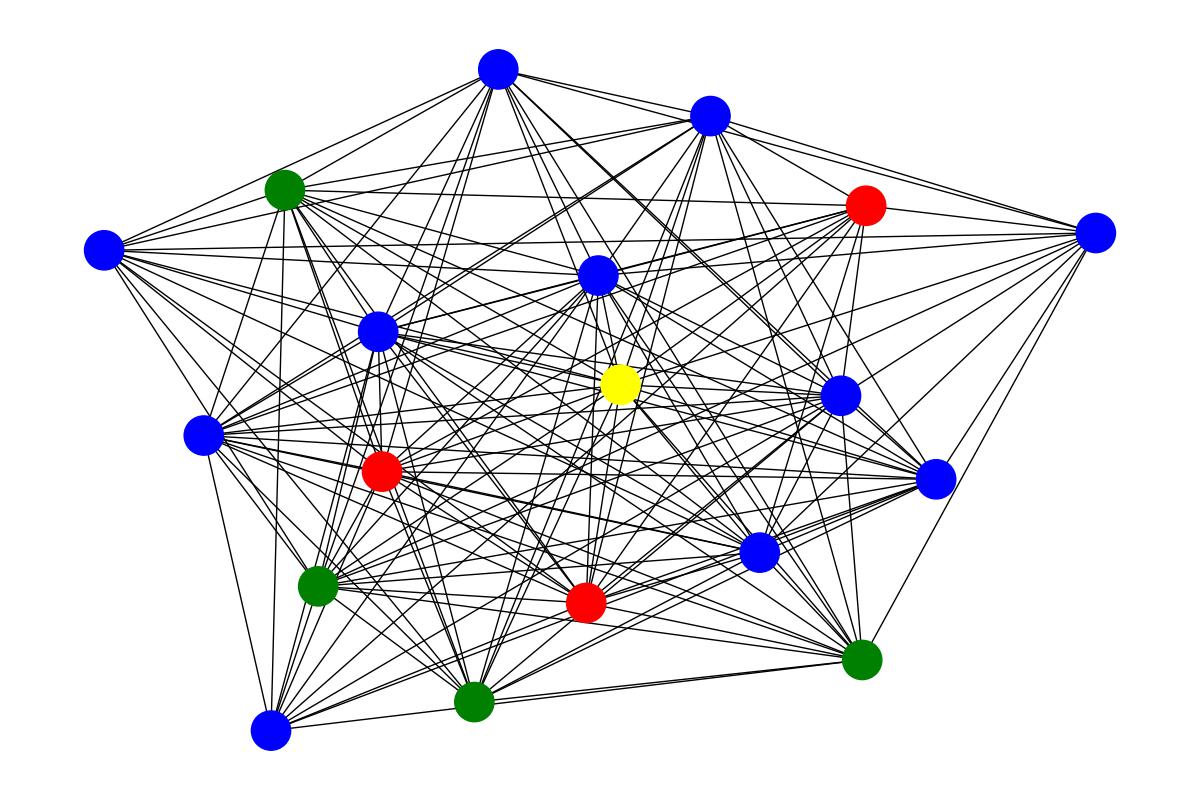}
        \label{fig:visual_tw_gsm}
    }\hfill
    \subfigure[Generation of EDGE.]{%
        \centering
        \includegraphics[width=0.3\textwidth]{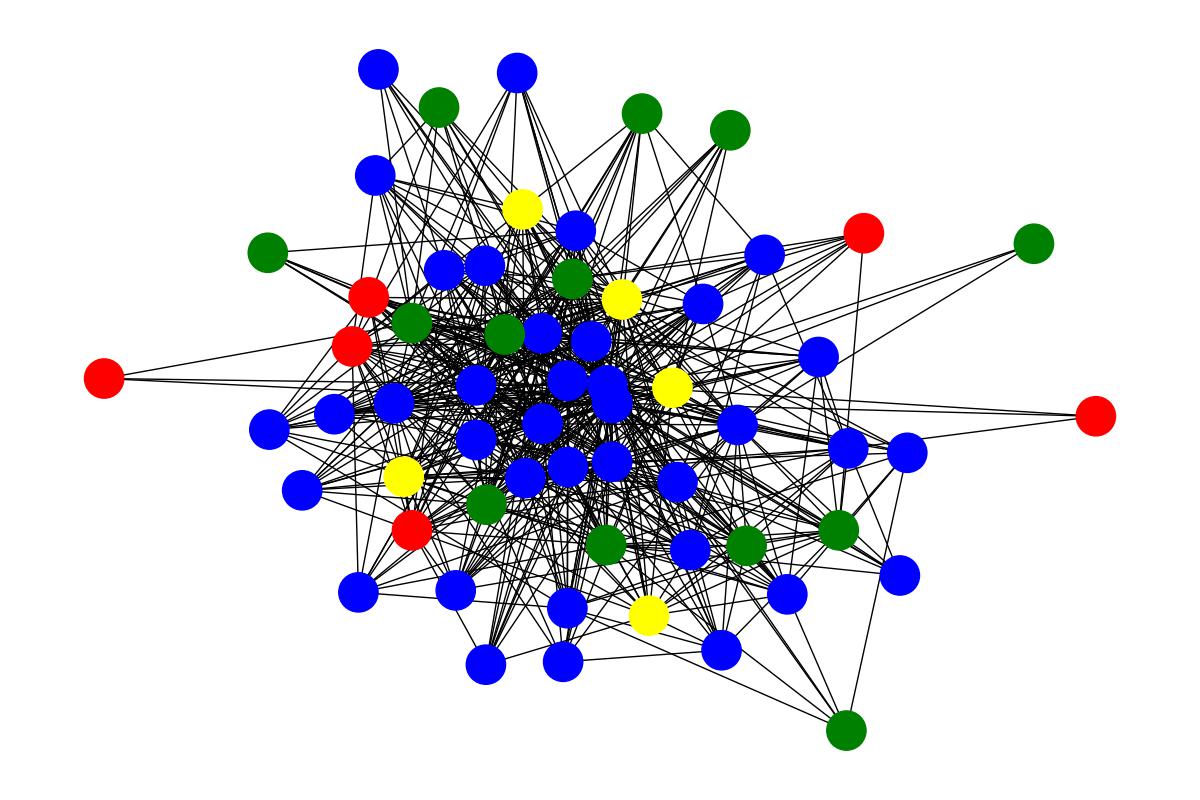}
        \label{fig:visual_tw_edge}
    }\hfill
    \subfigure[Generation of DiGress.]{%
        \centering
        \includegraphics[width=0.3\textwidth]{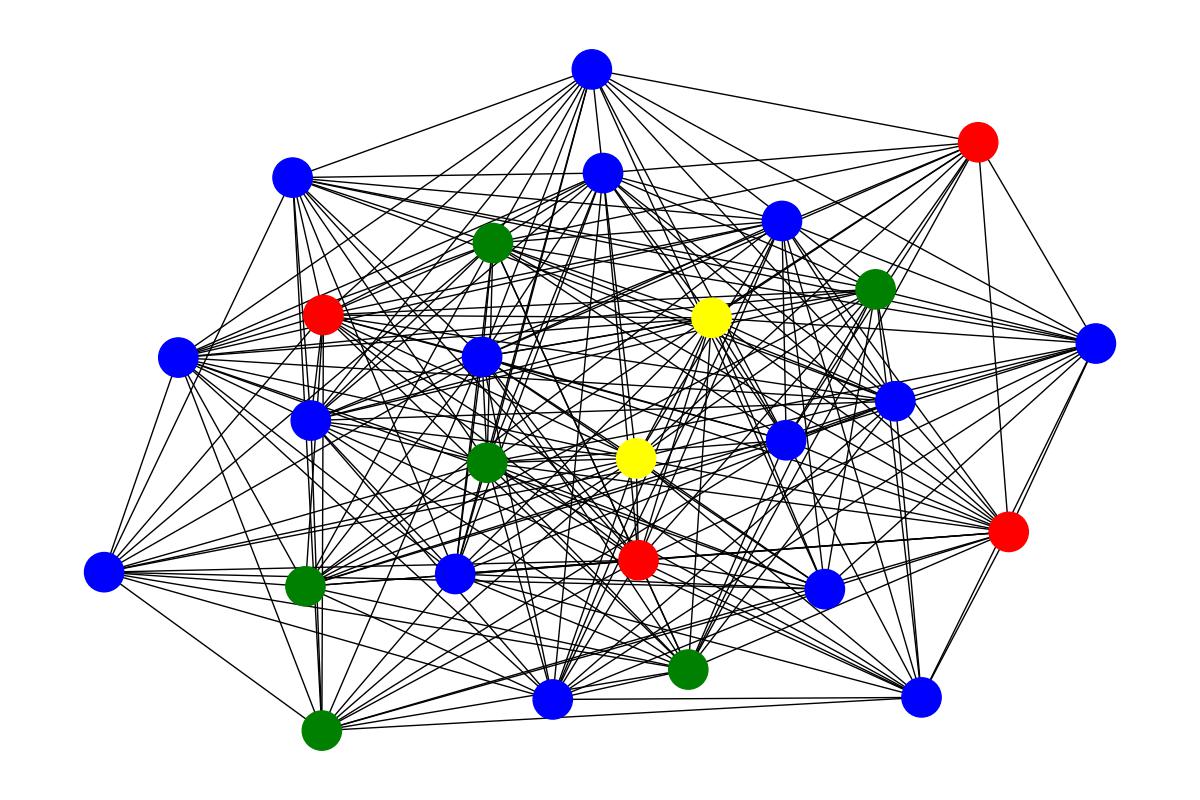}
        \label{fig:visual_tw_digress}
    }\hfill
    \subfigure[Generation of \algo.]{%
        \centering
        \includegraphics[width=0.3\textwidth]{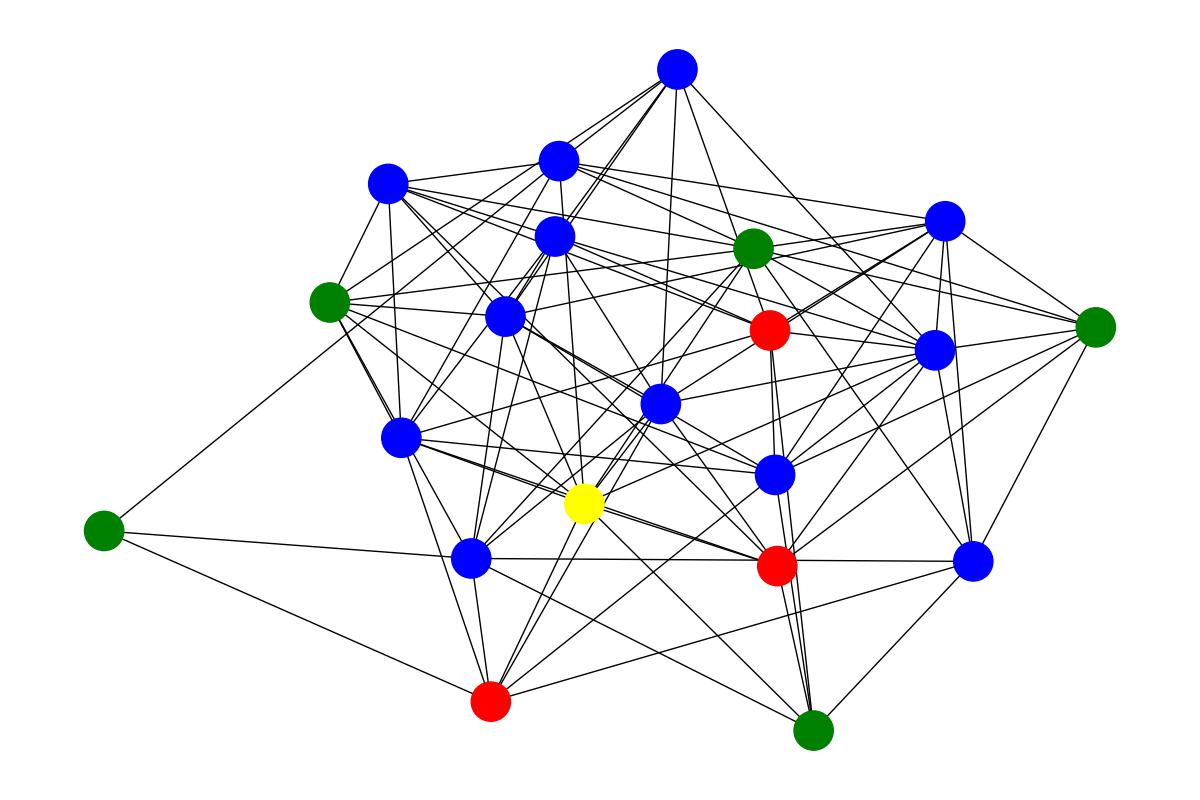}
        \label{fig:visual_tw_cdgraph}
    }\hfill%
     \caption{Visualization results of Twitter.}
     \label{fig:visual_tw}
\end{figure}

\subsection{Feasibility on social graph generation} 
In this section, we examine the dual conditions in two cases: those that are weakly positively correlated and those that are weakly negatively correlated. Additional cases are presented in Appendix~\ref{apdx:add_exp} due to space constraints. Note that since the conditions in Twitter and Flickr are all positively correlated, only the results from Facebook and BlogCatalog are reported for the case with weakly negatively correlated conditions.
Tables~\ref{tab:pos_summary} and~\ref{tab:neg_summary} demonstrate the feasibility of \algo\ and baselines on social graph generation in terms of validity, relative error ratios, and MMD for the cases with weakly positively correlated and weakly negatively correlated, respectively, conditions.

\textbf{Validity.}
Compared with the baselines, \algo\ is demonstrated to be able to generate graphs with the majority of nodes satisfying both of the specified conditions since it exploits the guidance of the dual condition classifier. Among the baselines, DiGress usually achieves the highest validity since it is aware of dual conditions, whereas SPECTRE, GSM, and EDGE do not account for them. Nevertheless, DiGress neglects the dependency between dual conditions and only achieves half of the validity achieved by \algo~.

\textbf{Relative error ratios of \#nodes, \#edges, and density, and MMD of clustering coefficients.}
\algo\ has superior performance in the relative error ratios of the edge number and density on Facebook and BlogCatalog for both cases of positive and negative correlations of specified conditions. Regarding the MMD of clustering coefficients, \algo\ demonstrates the best performance on Facebook and Twitter and also ranks second on Flickr and BlogCatalog. As the clustering coefficient is a fundamental property of social graphs, the results highlight that \algo\ surpasses the baselines in generating social graphs.

\textbf{Visualization}
To better understand the ability of social graph generation, we further present the visualization results. Figure~\ref{fig:visual_tw} visualizes the input and generated graphs for Twitter. Nodes in blue indicate that both specified conditions are met, while nodes in yellow and green represent those satisfying only one condition. Conversely, nodes in red do not meet any of the conditions. Obviously, the generation results of \algo\ are the closest to the input graph, thanks to the denoising process based on social homophily and social contagion. Due to the space constraint, additional visualization results are presented in Appendix~\ref{apdx:add_exp}.

\section{Conclusion}
In this paper, we make the first attempt to develop a conditional diffusion model for social networks, \algo~, which aims to synthesize social graphs satisfying two specified conditions. In contrast to previous studies on deep generative (diffusion) model-based graph generation, \algo~ introduces a novel feature of \textit{co-evolution dependency}. 
This feature integrates the concepts of \textit{social homophily} and \textit{social contagion}, allowing \algo\ to capture the interdependencies between specified conditions. Moreover, \algo\ exploits the \textit{dual-condition classifier} in the denoising process, ensuring that discrepancies in the diffusion process and structure reconstruction are jointly optimized. The experimental results manifest that \algo~achieves lower discrepancies in various network statistics under various correlations among specified conditions compared with the baselines.

%
%
%
\bibliographystyle{splncs04}
\bibliography{mybibliography}

\begin{thebibliography}{10}
\providecommand{\url}[1]{\texttt{#1}}
\providecommand{\urlprefix}{URL }
\providecommand{\doi}[1]{https://doi.org/#1}

\bibitem{CSUR2020Bonifati}
Bonifati, A., Holubov\'{a}, I., Prat-P\'{e}rez, A., Sakr, S.: Graph generators:
  State of the art and open challenges  \textbf{53}(2) (apr 2020).
  \doi{10.1145/3379445}, \url{https://doi.org/10.1145/3379445}

\bibitem{chen2023ICML}
Chen, X., He, J., Han, X., Liu, L.P.: Efficient and degree-guided graph
  generation via discrete diffusion modeling. In: International Conference on
  Machine Learning (2023)

\bibitem{NIPS2020Chenthamarakshan}
Chenthamarakshan, V., Das, P., Hoffman, S., Strobelt, H., Padhi, I., Lim, K.W.,
  Hoover, B., Manica, M., Born, J., Laino, T., Mojsilovic, A.: Cogmol:
  Target-specific and selective drug design for covid-19 using deep generative
  models. In: Larochelle, H., Ranzato, M., Hadsell, R., Balcan, M., Lin, H.
  (eds.) Advances in Neural Information Processing Systems. vol.~33, pp.
  4320--4332. Curran Associates, Inc. (2020),
  \url{https://proceedings.neurips.cc/paper_files/paper/2020/file/2d16ad1968844a4300e9a490588ff9f8-Paper.pdf}

\bibitem{SIGMOD2015Erling}
Erling, O., Averbuch, A., Larriba-Pey, J., Chafi, H., Gubichev, A., Prat, A.,
  Pham, M.D., Boncz, P.: The ldbc social network benchmark: Interactive
  workload. p. 619–630. SIGMOD '15, Association for Computing Machinery, New
  York, NY, USA (2015). \doi{10.1145/2723372.2742786}

\bibitem{CVPR2022Gu}
Gu, S., Chen, D., Bao, J., Wen, F., Zhang, B., Chen, D., Yuan, L., Guo, B.:
  Vector quantized diffusion model for text-to-image synthesis. In: Proceedings
  of the IEEE/CVF Conference on Computer Vision and Pattern Recognition (CVPR).
  pp. 10696--10706 (June 2022)

\bibitem{KDD2021Guo}
Guo, X., Du, Y., Zhao, L.: Deep generative models for spatial networks. In:
  Proceedings of the 27th ACM SIGKDD Conference on Knowledge Discovery \& Data
  Mining. p. 505–515. KDD '21, Association for Computing Machinery, New York,
  NY, USA (2021). \doi{10.1145/3447548.3467394},
  \url{https://doi.org/10.1145/3447548.3467394}

\bibitem{NIPS2020Ho}
Ho, J., Jain, A., Abbeel, P.: Denoising diffusion probabilistic models. In:
  Larochelle, H., Ranzato, M., Hadsell, R., Balcan, M., Lin, H. (eds.) Advances
  in Neural Information Processing Systems. vol.~33, pp. 6840--6851. Curran
  Associates, Inc. (2020)

\bibitem{NIPS2021hoogeboom}
Hoogeboom, E., Nielsen, D., Jaini, P., Forr{\'e}, P., Welling, M.: Argmax flows
  and multinomial diffusion: Learning categorical distributions. In:
  Beygelzimer, A., Dauphin, Y., Liang, P., Vaughan, J.W. (eds.) Advances in
  Neural Information Processing Systems (2021)

\bibitem{NIPS2022huang}
Huang, H., Sun, L., Du, B., Lv, W.: Conditional diffusion based on discrete
  graph structures for molecular graph generation. In: NeurIPS 2022 Workshop on
  Score-Based Methods (2022)

\bibitem{NIPS2019Liao}
Liao, R., Li, Y., Song, Y., Wang, S., Hamilton, W., Duvenaud, D.K., Urtasun,
  R., Zemel, R.: Efficient graph generation with graph recurrent attention
  networks. In: Wallach, H., Larochelle, H., Beygelzimer, A., d\textquotesingle
  Alch\'{e}-Buc, F., Fox, E., Garnett, R. (eds.) Advances in Neural Information
  Processing Systems. vol.~32. Curran Associates, Inc. (2019)

\bibitem{TCSS2020Luo}
Luo, W., Duan, B., Jiang, H., Ni, L.: Time-evolving social network generator
  based on modularity: Tesng-m. IEEE Transactions on Computational Social
  Systems  \textbf{7}(3),  610--620 (2020). \doi{10.1109/TCSS.2020.2979806}

\bibitem{pmlr2022martinkus}
Martinkus, K., Loukas, A., Perraudin, N., Wattenhofer, R.: {SPECTRE}: Spectral
  conditioning helps to overcome the expressivity limits of one-shot graph
  generators. In: Chaudhuri, K., Jegelka, S., Song, L., Szepesvari, C., Niu,
  G., Sabato, S. (eds.) Proceedings of the 39th International Conference on
  Machine Learning. Proceedings of Machine Learning Research, vol.~162, pp.
  15159--15179. PMLR (17--23 Jul 2022)

\bibitem{AISTATS2020niu}
Niu, C., Song, Y., Song, J., Zhao, S., Grover, A., Ermon, S.: Permutation
  invariant graph generation via score-based generative modeling. In:
  International Conference on Artificial Intelligence and Statistics. pp.
  4474--4484. PMLR (2020)

\bibitem{NIPS2022Saharia}
Saharia, C., Chan, W., Saxena, S., Li, L., Whang, J., Denton, E.L.,
  Ghasemipour, K., Gontijo~Lopes, R., Karagol~Ayan, B., Salimans, T., Ho, J.,
  Fleet, D.J., Norouzi, M.: Photorealistic text-to-image diffusion models with
  deep language understanding. In: Koyejo, S., Mohamed, S., Agarwal, A.,
  Belgrave, D., Cho, K., Oh, A. (eds.) Advances in Neural Information
  Processing Systems. vol.~35, pp. 36479--36494. Curran Associates, Inc. (2022)

\bibitem{JMLR2020Samanta}
Samanta, B., De, A., Jana, G., Gomez, V., Chattaraj, P.K., Ganguly, N.,
  Gomez-Rodriguez, M.: Nevae: A deep generative model for molecular graphs
  \textbf{21}(1) (jan 2020)

\bibitem{ICLR2022Savinov}
Savinov, N., Chung, J., Binkowski, M., Elsen, E., van~den Oord, A.:
  Step-unrolled denoising autoencoders for text generation. In: International
  Conference on Learning Representations (2022)

\bibitem{WWW2022Schweimer}
Schweimer, C., Gfrerer, C., Lugstein, F., Pape, D., Velimsky, J.A.,
  Els\"{a}sser, R., Geiger, B.C.: Generating simple directed social network
  graphs for information spreading. In: Proceedings of the ACM Web Conference
  2022. p. 1475–1485. WWW '22, Association for Computing Machinery, New York,
  NY, USA (2022). \doi{10.1145/3485447.3512194},
  \url{https://doi.org/10.1145/3485447.3512194}

\bibitem{Shi2020ICLR}
Shi, C., Xu, M., Zhu, Z., Zhang, W., Zhang, M., Tang, J.: Graphaf: a flow-based
  autoregressive model for molecular graph generation. In: International
  Conference on Learning Representations (2020)

\bibitem{TBD2018Shuai}
Shuai, H.H., Yang, D.N., Shen, C.Y., Yu, P.S., Chen, M.S.: Qmsampler: Joint
  sampling of multiple networks with quality guarantee. IEEE Transactions on
  Big Data  \textbf{4}(1),  90--104 (2018). \doi{10.1109/TBDATA.2017.2715847}

\bibitem{ICDM2013Shuai}
Shuai, H.H., Yang, D.N., Yu, P.S., Shen, C.Y., Chen, M.S.: On pattern
  preserving graph generation. In: 2013 IEEE 13th International Conference on
  Data Mining. pp. 677--686 (2013). \doi{10.1109/ICDM.2013.14}

\bibitem{ICLR2023vignac}
Vignac, C., Krawczuk, I., Siraudin, A., Wang, B., Cevher, V., Frossard, P.:
  Digress: Discrete denoising diffusion for graph generation. In: The Eleventh
  International Conference on Learning Representations (2023)

\bibitem{ICDE2021Wang}
Wang, C., Wang, B., Huang, B., Song, S., Li, Z.: Fastsgg: Efficient social
  graph generation using a degree distribution generation model. In: 2021 IEEE
  37th International Conference on Data Engineering (ICDE). pp. 564--575
  (2021). \doi{10.1109/ICDE51399.2021.00055}

\end{thebibliography}
%





\clearpage
\appendix
\section{Detailed derivations of the forward processes}\label{apdx:diffusion_process_derivation}
As mentioned in Sec. 3, each step of the forward noising processes of $\mathbf{X}_{c_{i}}$ (for each of the specified conditions $c_{i}, i \in \{1,2\}$) and $\mathbf{E}$ can be derived as follows: 
\begin{align}
    \notag&q(\mathbf{X}_{c_{i}}^{(t)}|\mathbf{X}_{c_{i}}^{(t-1)})=\mathbf{X}_{c_{i}}^{(t-1)}\mathbf{Q}_{X_{c_{i}}}^{(t)};q(\mathbf{E}^{(t)}|\mathbf{E}^{(t-1)})=\mathbf{E}^{(t-1)}\mathbf{Q}_{E_{c_{i}}}^{(t)},
\end{align}
where the transition matrices $\mathbf{Q}_{X_{c_{i}}}^{(t)},\mathbf{Q}_{E_{c_{i}}}^{(t)} \in \mathbb{R}^{2 \times 2}$ are defined with a parameter $\beta_{t} \in (0,1)$ as follows:
\begin{align*}
    \mathbf{Q}_{X_{c_{i}}}^{(t)}=
    \begin{bmatrix}
    1-\beta_{t} & \beta_{t}\\
    \beta_{t} & 1-\beta_{t}
    \end{bmatrix};
    \mathbf{Q}_{E_{c_{i}}}^{(t)}=
    \begin{bmatrix}
    1-\beta_{t} & \beta_{t}\\
    \beta_{t} & 1-\beta_{t}
    \end{bmatrix}.
\end{align*}
Note that it is equivalent to express $q(\mathbf{X}_{c_{i}}^{(t)}|\mathbf{X}_{c_{i}}^{(t-1)})$ and $q(\mathbf{E}^{(t)}|\mathbf{E}^{(t-1)})$ in the form of the Bernoulli distribution $\mathcal{B}$ as follows:
\begin{align}
    \notag q(\mathbf{X}_{c_{i}}^{(t)}|\mathbf{X}_{c_{i}}^{(t-1)})&=\mathcal{B}(\mathbf{X}_{c_{i}}^{(t)};(1-\beta_{t})\mathbf{X}_{c_{i}}^{(t-1)}+\beta_{t}\mathbf{1}/2);\\
    q(\mathbf{E}^{(t)}|\mathbf{E}^{(t-1)})&=\mathcal{B}(\mathbf{E}^{(t)};(1-\beta_{t})\mathbf{E}^{(t-1)}+\beta_{t}\mathbf{1}/2).
\end{align}

Recall that the overall forward processes of $\mathbf{X}_{c_i}$ and $\mathbf{E}$ are formulated as follows:
\begin{align}
    \notag q(\mathbf{X}_{c_{i}}^{(0:T)})&=\prod_{t=1}^{T}q(\mathbf{X}_{c_{i}}^{(t)}|\mathbf{X}_{c_{i}}^{(t-1)},\mathbf{E}^{(t-1)})q(\mathbf{E}^{(t-1)}|\mathbf{X}_{c_{i}}^{(t-1)}),\\
    q(\mathbf{E}^{(0:T)})&=\prod_{t=1}^{T}q(\mathbf{E}^{(t)}|\mathbf{E}^{(t-1)},\mathbf{X}_{c_{i}}^{(t-1)})q(\mathbf{X}_{c_{i}}^{(t-1)}|\mathbf{E}^{(t-1)}),
\end{align}
where the dependency between $\mathbf{X}_{c_{i}}^{(t-1)}$ and $\mathbf{E}^{(t-1)}$ can be determined by the properties of social homophily and social contagion as follows. 

For social contagion, given the existence of an edge $e_{m,n}$ (i.e., $\mathbf{e}_{m,n}^{(t-1)}=1$), we define the probability whether node $x_{m}$ satisfies condition $c_{i}$ (i.e., $\mathbf{x}_{c_{i},m}^{(t-1)}=1$) or not as follows.
\begin{align}
    q(\mathbf{x}_{c_{i},m}^{(t-1)}|\mathbf{e}_{m,n}^{(t-1)}=1,\mathbf{x}_{c_{i},n}^{(t-1)})=
    \begin{cases}
        p, \mathbf{x}_{c_{i},m}^{(t-1)}=\mathbf{x}_{c_{i},n}^{(t-1)},\\
        1-p, \mathbf{x}_{c_{i},m}^{(t-1)} \neq \mathbf{x}_{c_{i},n}^{(t-1)}.
    \end{cases}
\end{align}
For social homophily, we define the probability of the existence of an edge $\mathbf{e}_{m,n}$ from given node condition $\mathbf{x}_{c_{i},m}^{(t-1)}$ by marginalization with respect to another node condition $\mathbf{x}_{c_{i},n}^{(t-1)}$ as follows:
\begin{align}
    q(\mathbf{e}_{m,n}^{(t-1)}=1|\mathbf{x}_{c_{i},m}^{(t-1)})=
    \sum_{\mathbf{x}_{c_{i},n}^{(t-1)} \in \{0,1\}}q(\mathbf{e}_{m,n}^{(t-1)}=1|\mathbf{x}_{c_{i},m}^{(t-1)},\mathbf{x}_{c_{i},n}^{(t-1)})q(\mathbf{x}_{c_{i},n}^{(t-1)}|\mathbf{x}_{c_{i},m}^{(t-1)}),
\end{align}
where $q(\mathbf{x}_{c_{i},n}^{(t-1)}|\mathbf{x}_{c_{i},m}^{(t-1)})$ can be computed with the bivariate Bernoulli distribution of $\mathbf{x}_{c_{i},n}^{(t-1)}$ and $\mathbf{x}_{c_{i},m}^{(t-1)}$, and
\begin{align*}
    q(\mathbf{e}_{m,n}^{(t-1)}=1|\mathbf{x}_{c_{i},m}^{(t-1)},\mathbf{x}_{c_{i},n}^{(t-1)})=
    \begin{cases}
        1, \mathbf{x}_{c_{i},m}^{(t-1)}=\mathbf{x}_{c_{i},n}^{(t-1)},\\
        0, \mathbf{x}_{c_{i},m}^{(t-1)} \neq \mathbf{x}_{c_{i},n}^{(t-1)}.
    \end{cases}
\end{align*}

\section{Pseudocode of the proposed algorithms}\label{apdx:pseudocode}
The training procedure of \algo~is depicted in Algorithm~\ref{alg:training_CDGraph}.
\begin{algorithm}
\caption{\algo~Training}\label{alg:training_CDGraph}
\begin{algorithmic}
\Require A condition indication graph $G_{C}=(\{\mathbf{X}_{c_i}\}_{i=1}^{|C|},\mathbf{E})$ with condition set $C=\{c_{1},c_{2}\}$.
\State Sample $t \sim \{1,2,\cdots,T\}$
\State Sample $G_{C}^{(t)} \sim (\{\mathbf{X}_{c_{i}}\overline{\mathbf{Q}}_{X_{c_{i}}}^{(t)}\}_{c_i \in C},\mathbf{E}\overline{\mathbf{Q}}_{E}^{(t)})$
\State Extract structural and spectral features $z \longleftarrow f(G_{C}^{t},t)$
\State $\hat{p}_{X_{c_{i}}}, \hat{p}_{E} \longleftarrow \phi_{\theta}(G_{C}^{t}, z)$
\State Optimize the cross-entropy loss $l_{CE}(\hat{p}_{X_{c_{1}}},\mathbf{X}_{c_{1}})+l_{CE}(\hat{p}_{X_{c_{2}}},\mathbf{X}_{c_{2}})+\lambda l_{CE}(\hat{p}_{E},\mathbf{E})$
\end{algorithmic}
\end{algorithm}

The sampling procedure of \algo~is depicted in Algorithm~\ref{alg:sampling_CDGraph}.
\begin{algorithm}
\caption{Sampling from \algo}\label{alg:sampling_CDGraph}
\begin{algorithmic}
\State Sample $n$ nodes from training data distribution
\State Sample $G_{C}^{T} \sim (\{q(\mathbf{X}_{c_{i}}^{(T)})\}_{c_{i} \in C},q(\mathbf{E}^{(T)}))$
\For{$t=T$ to $1$}
\State Extract structural and spectral features $z \longleftarrow f(G_{C}^{t},t)$
\State $\hat{p}_{X_{c_{1}}}, \hat{p}_{E} \longleftarrow \phi_{\theta}(G_{C}^{t}, z)$
\State $p_{\theta}(x_{n,c_{i}}^{(t-1)}|x_{n,c_{i}}^{(t)},e_{n,m}^{(t)}) \longleftarrow \sum_{x_{n,c_{i}}^{(0)} \in \{0,1\}}q(x_{n,c_{i}}^{(t-1)}|x_{n,c_{i}}^{(t)},x_{n,c_{i}}^{(0)})\hat{p}_{X_{c_{i}}}(x_{n,c_{i}}^{(0)}|x_{n,c_{i}}^{(t)},e_{n,m}^{(t)})$ for $c_{i} \in C$.
\State $p_{\theta}(e_{n,m}^{(t-1)}|e_{n,m}^{(t)},x_{n,C}^{(t)},x_{m,C}^{(t)}) \longleftarrow \sum_{e_{n,m}^{(0)} \in \{0,1\}}q(e_{n,m}^{(t-1)}|e_{n,m}^{(t)},e_{n,m}^{(0)})\hat{p}_{E}(e_{n,m}^{(0)}|e_{n,m}^{(t)},x_{n,C}^{(t)},x_{m,C}^{(t)})$
\State Sample $\mathbf{X}_{c_{i}}^{(t-1)} \sim \prod_{n=1}^{N}\prod_{(n,m) \in E}p_{\theta}(x_{n,c_{i}}^{(t-1)}|x_{n,c_{i}}^{(t)},e_{n,m}^{(t)})$
\State Sample $\mathbf{E}^{(t-1)} \sim \prod_{n=1}^{N}\prod_{(n,m) \in E}p_{\theta}(e_{n,m}^{(t-1)}|e_{n,m}^{(t)},x_{n,C}^{(t)},x_{m,C}^{(t)})$
\EndFor
\Ensure $\hat{G}_{C}=(\{\hat{\mathbf{X}}_{c_i}\}_{c_{i} \in C},\hat{\mathbf{E}})$
\end{algorithmic}
\end{algorithm}

The pseudocode of the sampling of the diffusion process is shown in the Algorithm~\ref{alg:condition_diffusion}.
\begin{algorithm}
\caption{Conditional Sampling}\label{alg:condition_diffusion}
\begin{algorithmic}
\Require A condition indication graph $G_{C}=(\{\mathbf{X}_{c_i}\}_{i=1}^{|C|},\mathbf{E})$ with condition set $C$.
\State Sample $G_{C}^{T} \sim (q(\mathbf{X}_{c_{i}}^{(T)}),q(\mathbf{E}^{(T)}))$
\For{$t=T$ to $1$}
\State $\hat{p}_{c_{i}},\hat{p}_{e} \leftarrow \phi_{\theta}(G_{C}^{(t)})$
\State $\hat{c}_{i} \leftarrow g_{c_{i}}(G_{C}^{(t)})$
\State $\hat{c}_{j} \leftarrow g_{c_{j}}(G_{C}^{(t)})$ for $G_{C}^{(t)}$ with $\hat{c}_{i}=1$.
\State $G_{C}^{(t-1)} \sim p_{c_{j}}(\hat{c_{j}}|G_{C}^{(t-1)},c_{i})p_{c_{i}}(\hat{c_{i}}|G_{C}^{(t-1)})p_{\theta}(G_{C}^{(t-1)}|G_{C}^{(t)})$
\EndFor
\Ensure $\hat{G}_{C}=(\{\hat{\mathbf{X}}_{c_i}\}_{c_{i} \in C},\hat{\mathbf{E}})$
\end{algorithmic}
\end{algorithm}

\section{Additional Experimental Results}\label{apdx:add_exp}
\subsection{More evaluation results}
We present the evaluation results under medium/high positive and negative correlations in Tables~\ref{tab:med_pos_summary} to~\ref{tab:high_neg_summary}.
\begin{table}[hbtp]
    \centering
    \scriptsize
    \begin{tabular}{l | c c c c c | c c c c c }
        \hline
        Dataset & \multicolumn{5}{c|}{Facebook} & \multicolumn{5}{c}{BlogCatalog} \\ \hline
        Metric & Validity & Node & Edge & Density & Clust. coeff. & Validity & Node & Edge & Density & Clust. coeff. \\\hline
        SPECTRE & \underline{0.379} & \underline{0.084} & 1.484 & 1.021 & 0.534 & 0.38 & \underline{0.004} & 5.621 & 5.135 & 0.099  \\\hline
        GSM & 0.33 & \textbf{0.083} & \underline{0.445} & 0.360 & 0.03 & 0.289 & \textbf{0.002} & 3.398 & 3.404 & 0.072  \\\hline
        EDGE & 0.307 & 0.919 & 2.528 & 0.934 & \underline{0.029} & 0.282 & 0.420 & \underline{0.562} & \underline{0.323} & \textbf{0.013}  \\\hline
        DiGress & 0.188 & 0.341 & 0.446 & \underline{0.097} & 0.49 & 0.1875 & 0.063 & 0.779 & 0.892 & 0.342  \\\hline
        CDGraph & \textbf{1} & 0.149 & \textbf{0.186} & \textbf{0.002} & \textbf{0.022} & \textbf{1} & 0.020 & \textbf{0.259} & \textbf{0.219} & \underline{0.031} \\\hline\hline
        Dataset & \multicolumn{5}{c|}{Twitter} & \multicolumn{5}{c}{Flickr} \\ \hline
        Metric & Validity & Node & Edge & Density & Clust. coeff. & Validity & Node & Edge & Density & Clust. coeff. \\\hline
        SPECTRE & 0.304 & 0.239 & 1.125 & \textbf{0.001} & 0.515 & 0.226 & 0.025 & 2.294 & 2.373 & 1.006  \\\hline
        GSM & 0.294 & 0.212 & 0.404 & \underline{0.045} & \underline{0.011} & 0.288 & \textbf{0.007} & 1.055 & 1.033 & \textbf{0.013}  \\\hline
        EDGE & 0.299 & \underline{0.107} & \underline{0.042} & 0.304 & \textbf{0.002} & 0.265 & 0.427 & 0.799 & 0.681 & \underline{0.014}  \\\hline
        DiGress & \underline{0.375} & 0.165 & 0.071 & 0.284 & 0.257 & \underline{0.5} & 0.039 &\underline{0.129} & \underline{0.089} & 0.307  \\\hline
        CDGraph & \textbf{1} & \textbf{0.045} & \textbf{0.001} & 0.409 & 0.016 & \textbf{1} & \underline{0.008} & \textbf{0.056} & \textbf{0.068} & 0.068 \\\hline
    \end{tabular}
    \caption{Summarization of evaluation results ($0.125<\rho<0.25$)}
    \label{tab:med_pos_summary}
\end{table}

\begin{table}[hbtp]
    \centering
    \scriptsize
    \begin{tabular}{l | c c c c c | c c c c c }
        \hline
        Dataset & \multicolumn{5}{c|}{Facebook} & \multicolumn{5}{c}{BlogCatalog} \\ \hline
        Metric & Validity & Node & Edge & Density & Clust. coeff. & Validity & Node & Edge & Density & Clust. coeff. \\\hline
        SPECTRE & 0.276 & 0.292 & 2.341 & 1.190 & 0.217 & 0.26 & 0.018 & 5.800 & 5.304 & 0.096 \\\hline
        GSM & 0.298 & \textbf{0.090} & \underline{0.433} & 0.371 & 0.029 & 0.242 & \underline{0.024} & 3.562 & 3.472 & 0.078 \\\hline
        EDGE & 0.32 & 0.782 & 2.039 & 0.817 & \underline{0.024} & 0.26 & 0.411 & \underline{0.559} & \underline{0.334} & \textbf{0.012} \\\hline
        DiGress & \underline{0.5} & 0.470 & 0.664 & \underline{0.286} & 0.244 & \underline{0.313} & 0.041 & 0.910 & 0.936 & 0.349 \\\hline
        CDGraph & \textbf{1} & \underline{0.149} & \textbf{0.186} & \textbf{0.002} & \textbf{0.022} & \textbf{1} & \textbf{0.020} & \textbf{0.259} & \textbf{0.219} & \underline{0.031} \\\hline
    \end{tabular}
    \caption{Summarization of evaluation results ($-0.25<\rho<-0.125$)}
    \label{tab:med_neg_summary}
\end{table}

\begin{table}[hbtp]
    \centering
    \scriptsize
    \begin{tabular}{l | c c c c c | c c c c c }
        \hline
        Dataset & \multicolumn{5}{c|}{Facebook} & \multicolumn{5}{c}{BlogCatalog} \\ \hline
        Metric & Validity & Node & Edge & Density & Clust. coeff. & Validity & Node & Edge & Density & Clust. coeff. \\\hline
        SPECTRE & 0.276 & \underline{0.110} & 1.113 & 0.439 & 0.288 & 0.2 & 0.065 & 7.036 & 6.291 & 0.092 \\\hline
        GSM & 0.274 & 0.147 & \underline{0.677} & 0.457 & 0.031 & 0.26 & \textbf{0.015} & 3.499 & 3.446 & 0.075 \\\hline
        EDGE & 0.307 & 0.897 & 2.424 & 0.926 & \underline{0.03} & \underline{0.262} & 0.391 & \underline{0.496} & \underline{0.269} & \textbf{0.013} \\\hline
        DiGress & \underline{0.375} & \textbf{0.002} & 0.725 & \underline{0.297} & 0.259 & 0.188 & 0.037 & 0.887 & 0.946 & 0.269 \\\hline
        CDGraph & \textbf{1} & 0.149 & \textbf{0.186} & \textbf{0.002} & \textbf{0.022} & \textbf{1} & \underline{0.020} & \textbf{0.259} & \textbf{0.219} & \underline{0.031} \\\hline\hline
        Dataset & \multicolumn{5}{c|}{Twitter} & \multicolumn{5}{c}{Flickr} \\ \hline
        Metric & Validity & Node & Edge & Density & Clust. coeff. & Validity & Node & Edge & Density & Clust. coeff. \\\hline
        SPECTRE & \underline{0.37} & 0.294 & 0.450 & \underline{0.062} & 0.46 & 0.323 & 0.032 & 2.268 & 2.348 & 1.006 \\\hline
        GSM & 0.326 & 0.192 & 0.530 & \textbf{0.009} & \underline{0.012} & \underline{0.285} & \underline{0.009} & 1.065 & 1.036 & \textbf{0.014} \\\hline
        EDGE & 0.289 & \underline{0.111} & \underline{0.068} & 0.317 & \textbf{0.002} & 0.319 & 0.429 & 0.801 & 0.696 & \textbf{0.014} \\\hline
        DiGress & 0.188 & 0.348 & 0.514 & 0.422 & 0.15 & 0.438 & \underline{0.011} & \textbf{0.023} & \textbf{0.040} & 0.371 \\\hline
        CDGraph & \textbf{1} & \textbf{0.045} & \textbf{0.001} & 0.409 & 0.016 & \textbf{1} & \textbf{0.008} & \underline{0.056} & \underline{0.068} & \underline{0.068} \\\hline
    \end{tabular}
    \caption{Summarization of evaluation results ($0.25<\rho<0.5$)}
    \label{tab:high_pos_summary}
\end{table}

\begin{table}[hbtp]
    \centering
    \scriptsize
    \begin{tabular}{l | c c c c c | c c c c c }
        \hline
        Dataset & \multicolumn{5}{c|}{Facebook} & \multicolumn{5}{c}{BlogCatalog} \\ \hline
        Metric & Validity & Node & Edge & Density & Clust. coeff. & Validity & Node & Edge & Density & Clust. coeff. \\\hline
        SPECTRE & 0.172 & 0.345 & \underline{0.312} & \underline{0.015} & 0.1376 & 0.22 & 0.027 & 4.879 & 4.521 & 0.087 \\\hline
        GSM & \underline{0.298} & \textbf{0.076} & 0.350 & 0.337 & 0.032 & 0.225 & \textbf{0.014} & 3.487 & 3.443 & 0.074 \\\hline
        EDGE & 0.294 & 0.817 & 2.138 & 0.873 & 0.023 & 0.235 & 0.405 & 0.544 & \underline{0.315} & \textbf{0.013} \\\hline
        DiGress & 0.25 & 0.463 & 0.653 & 0.266 & \textbf{0.19} & \underline{0.25} & 0.133 & \underline{0.49} & 0.710 & 0.209 \\\hline
        CDGraph & \textbf{1} & \underline{0.149} & \textbf{0.186} & \textbf{0.002} & \underline{0.022} & \textbf{1} & \underline{0.02} & \textbf{0.259} & \textbf{0.219} & \underline{0.031} \\\hline
    \end{tabular}
    \caption{Summarization of evaluation results ($-0.5<\rho<-0.25$)}
    \label{tab:high_neg_summary}
\end{table}

\subsection{Sensitivity tests}
We conduct a sensitivity tests to evaluate the validity of the generated graphs under various degrees of negative correlations $\rho$ of specified conditions. The degrees of negative correlation are low ($-0.125 < \rho < 0$), medium ($-0.25 < \rho < -0.125$) and high ($-0.5 < \rho < -0.25$).

We evaluate the validity under various degrees of negative correlation between specified conditions on Facebook and BlogCatalog in Figs.~\ref{fig:valid_sen_FB} and~\ref{fig:valid_sen_BC}, respectively. In both datasets, \algo~is able to achieve higher validity than all baselines under different degrees of negative correlation since it captures the dependencies between nodes with social contagion in the denoising process.
\begin{figure}
    \centering
    \subfigure[Facebook.]{%
        \centering
        \includegraphics[width=0.5\textwidth]{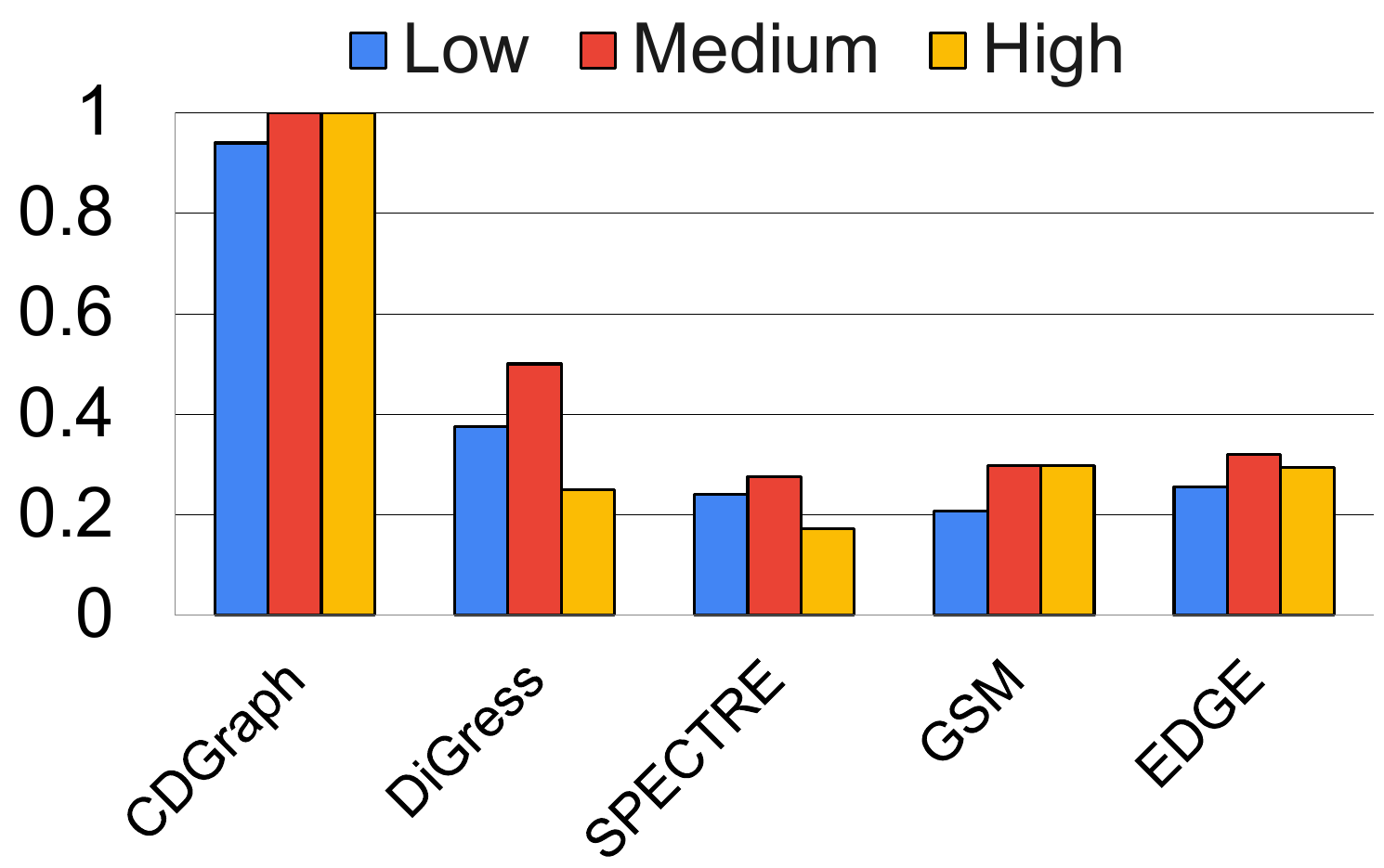}
        \label{fig:valid_sen_FB}
    }%
    \subfigure[BlogCatalog.]{%
        \centering
        \includegraphics[width=0.5\textwidth]{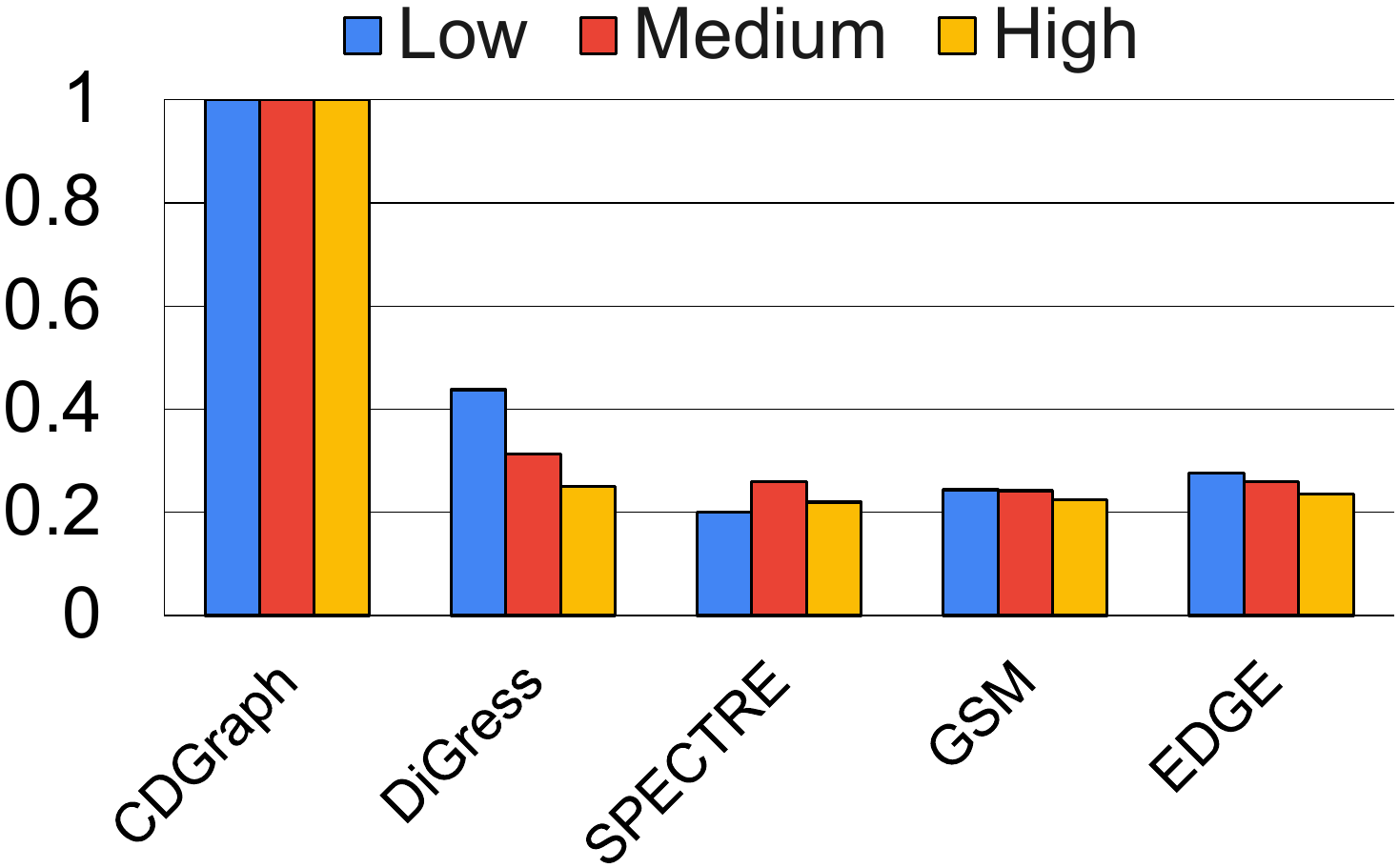}
        \label{fig:valid_sen_BC}
    }
     \caption{Validity under various degrees of negative correlation.}
    \label{fig:valid_sen}
\end{figure}

We evaluate the MMD of clustering coefficient under various degrees of negative correlation between specified conditions on Facebook and BlogCatalog in Figs.~\ref{fig:mmd_cluster_sen_FB} and~\ref{fig:mmd_cluster_sen_BC}, respectively. In both datasets, \algo~is able to generate social graphs with lower MMD of clustering coefficient since the connectivity between nodes can be effectively recovered due to social homophily-based co-evolution in the denoising process.
\begin{figure}
    \centering
    \subfigure[Facebook.]{%
        \centering
        \includegraphics[width=0.5\textwidth]{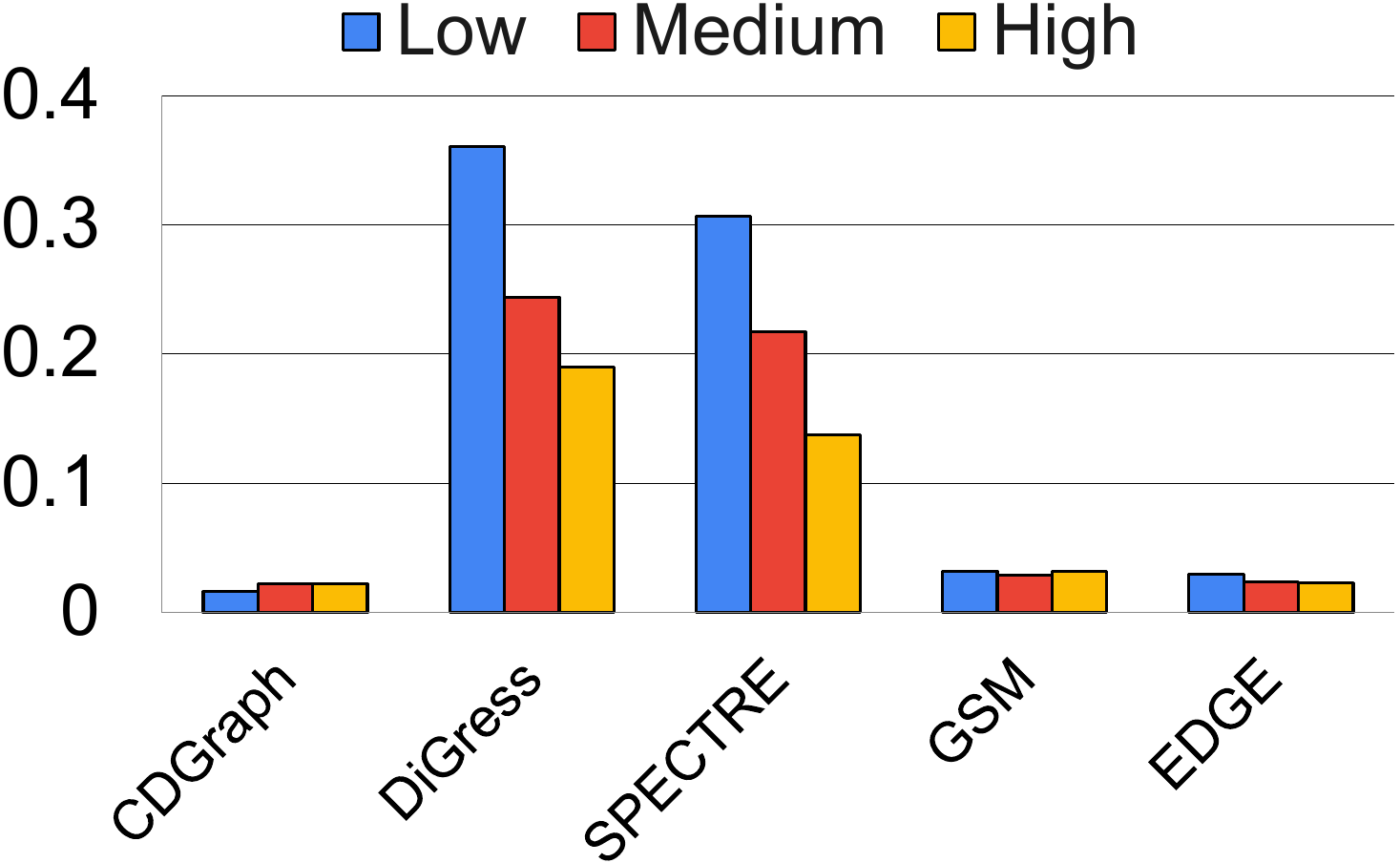}
        \label{fig:mmd_cluster_sen_FB}
    }%
    \subfigure[BlogCatalog.]{%
        \centering
        \includegraphics[width=0.5\textwidth]{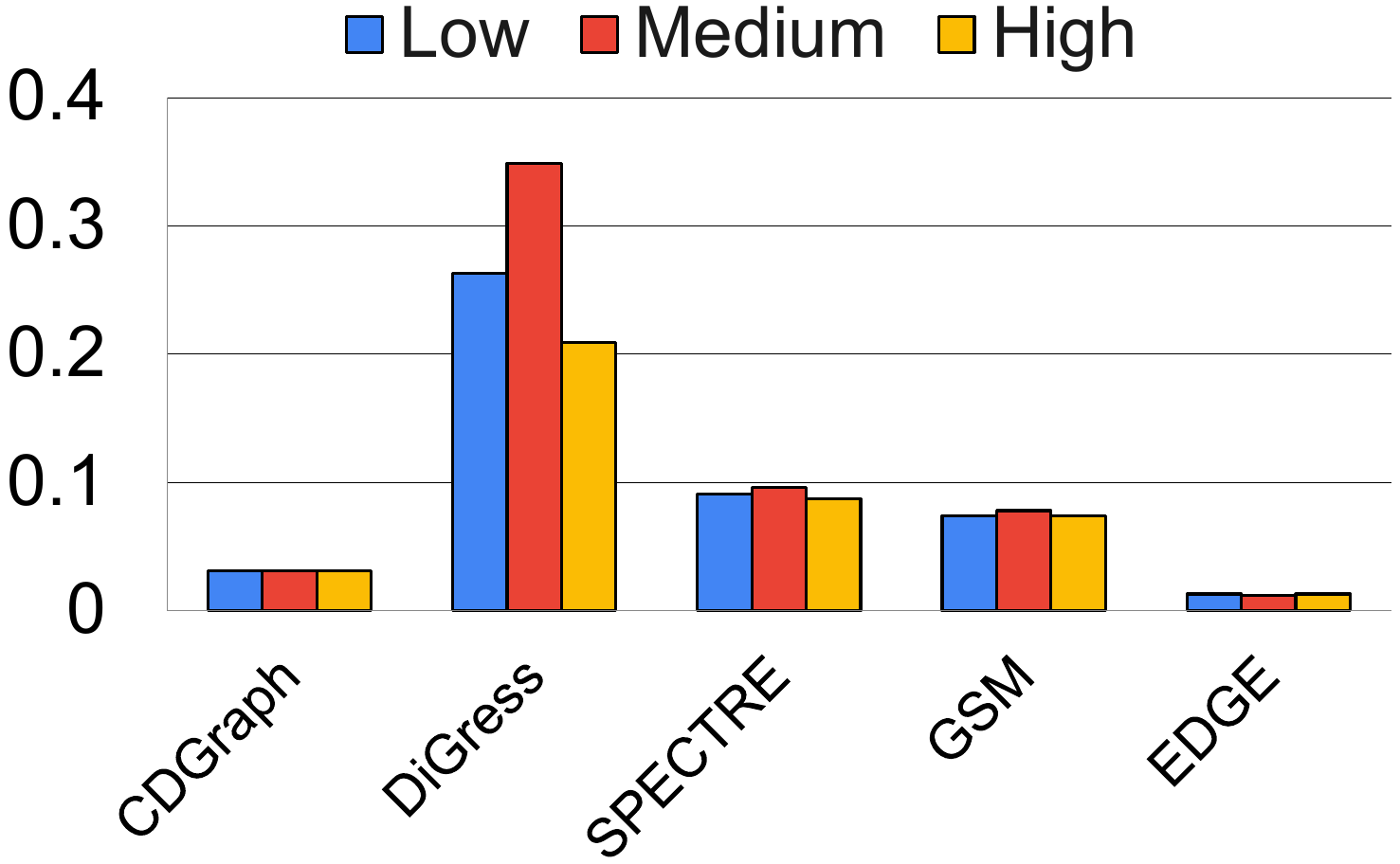}
        \label{fig:mmd_cluster_sen_BC}
    }
     \caption{MMD of clustering coefficient under various degrees of negative correlation.}
    \label{fig:mmd_cluster_sen}
\end{figure}

\subsection{Visualization}
We present the visualization results of Facebook in Figs.~\ref{fig:visual_fb}. The nodes in green represent that both specified conditions are satisfied.
\begin{figure}
    \centering
    \subfigure[Input graph.]{%
        \centering
        \includegraphics[width=0.3\textwidth]{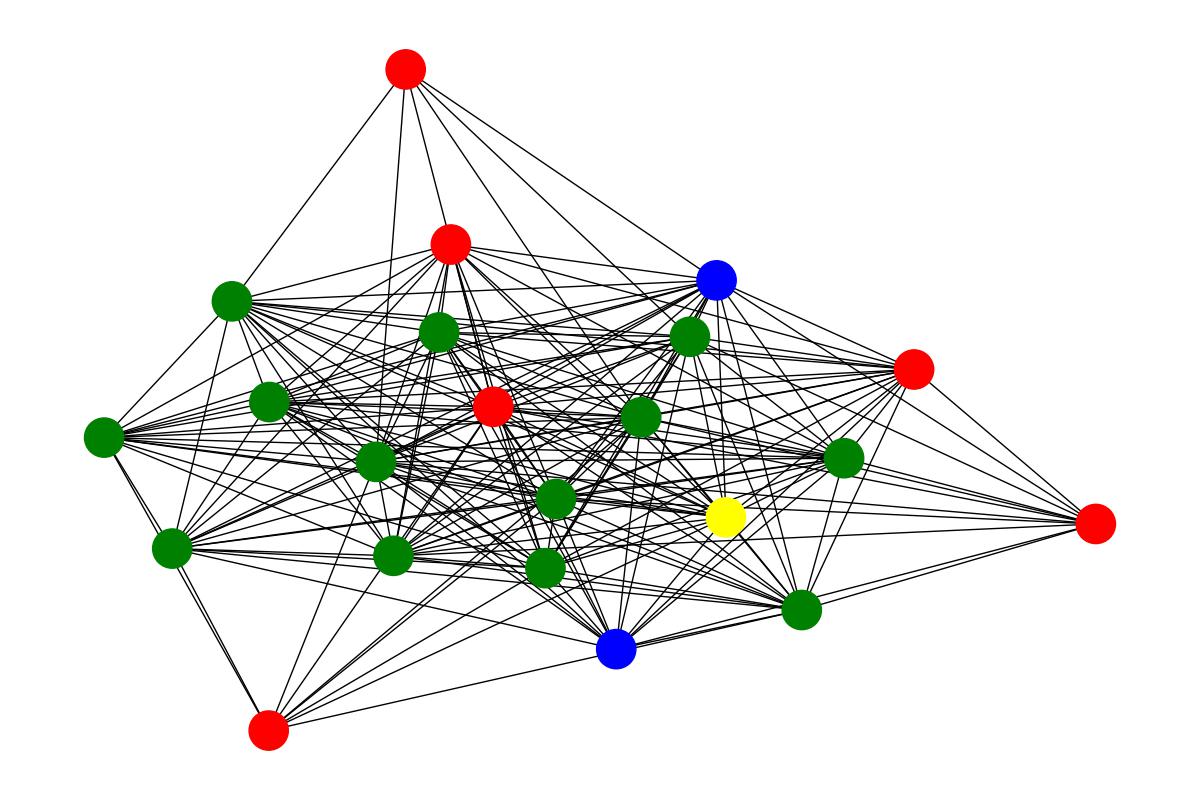}
        \label{fig:visual_fb_input}
    }\hfill
    \subfigure[Generation of SPECTRE.]{%
        \centering
        \includegraphics[width=0.3\textwidth]{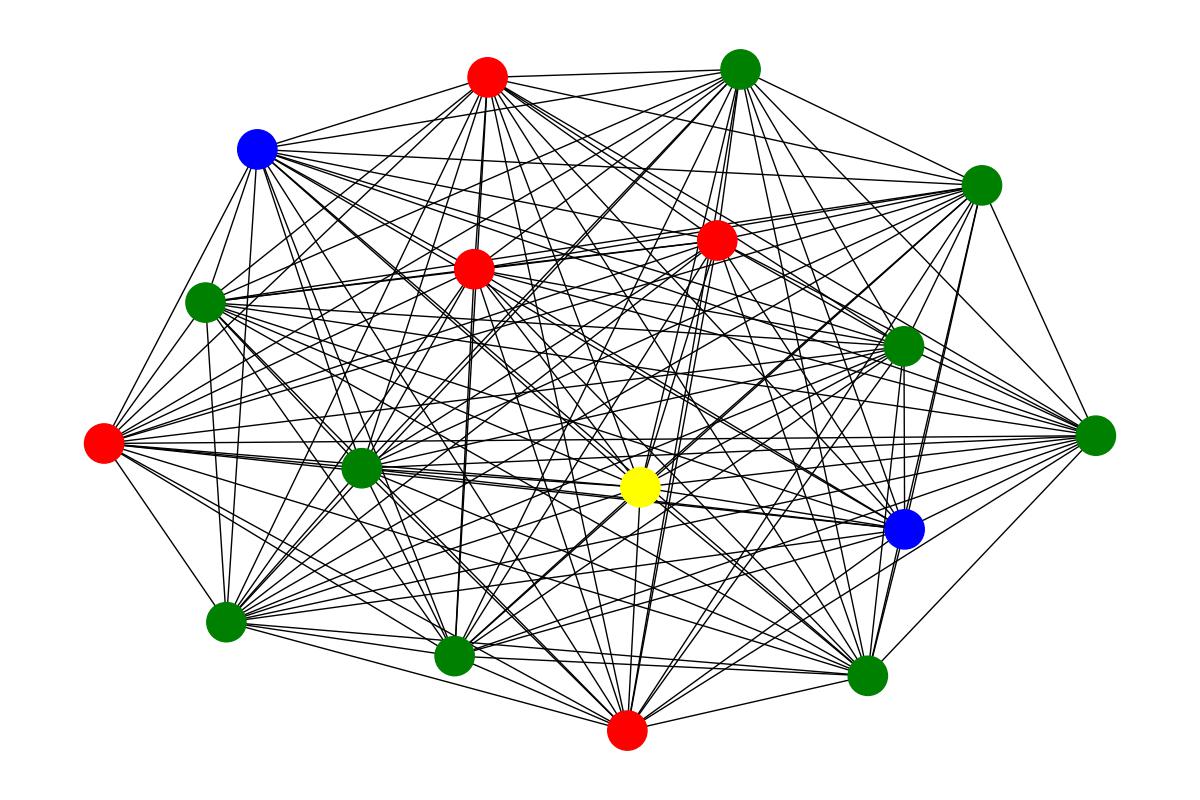}
        \label{fig:visual_fb_spectre}
    }\hfill
    \subfigure[Generation of GSM.]{%
        \centering
        \includegraphics[width=0.3\textwidth]{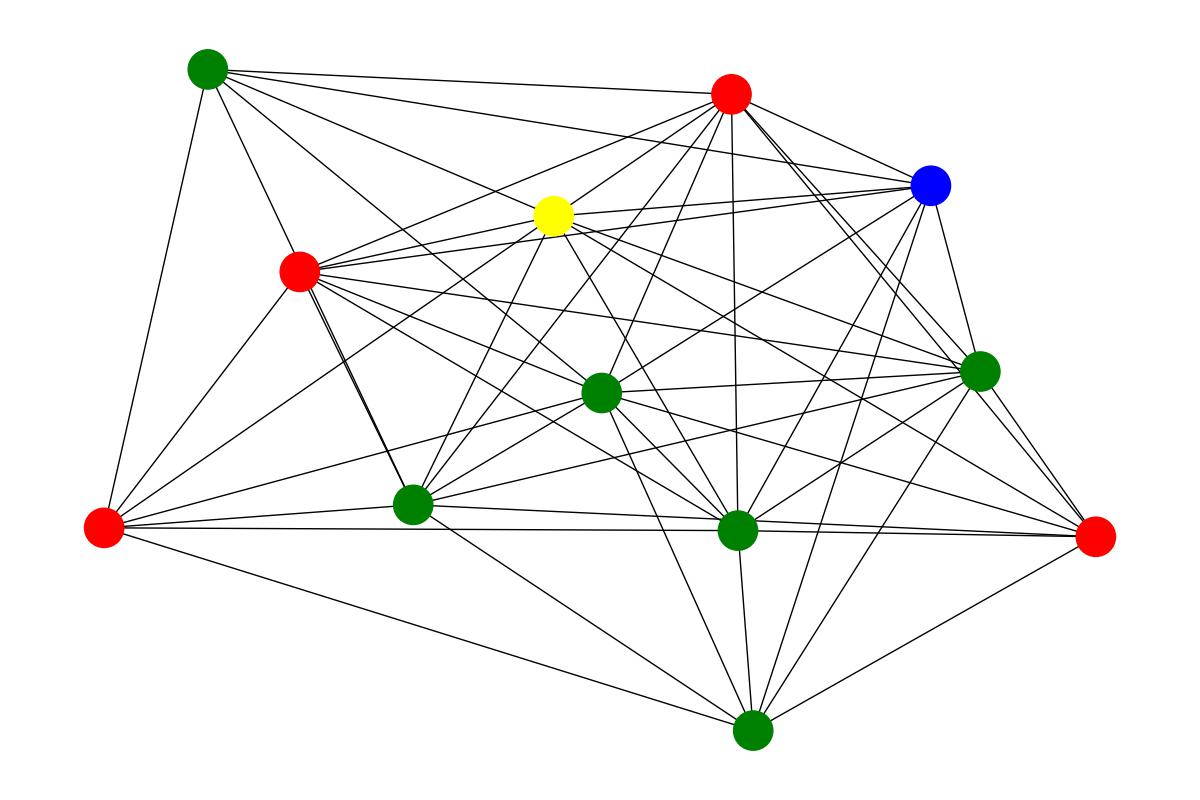}
        \label{fig:visual_fb_gsm}
    }\hfill
    \subfigure[Generation of EDGE.]{%
        \centering
        \includegraphics[width=0.3\textwidth]{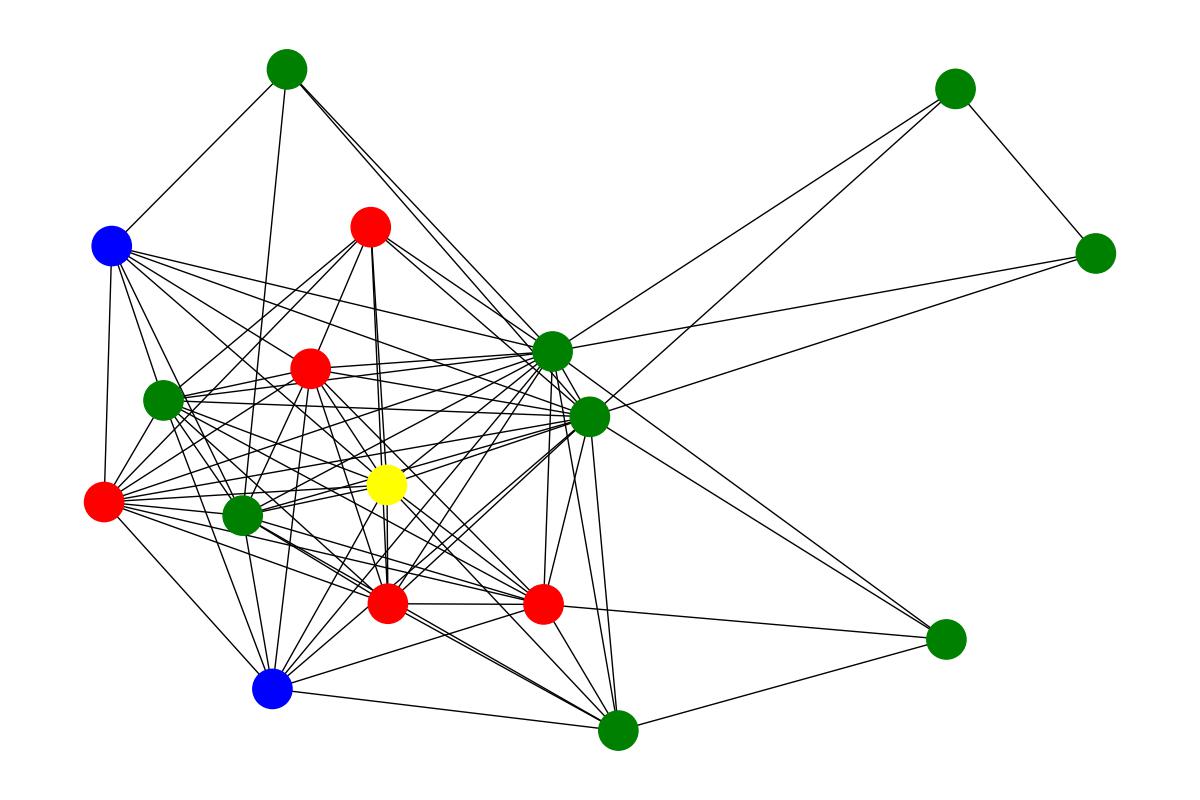}
        \label{fig:visual_fb_edge}
    }\hfill
    \subfigure[Generation of DiGress.]{%
        \centering
        \includegraphics[width=0.3\textwidth]{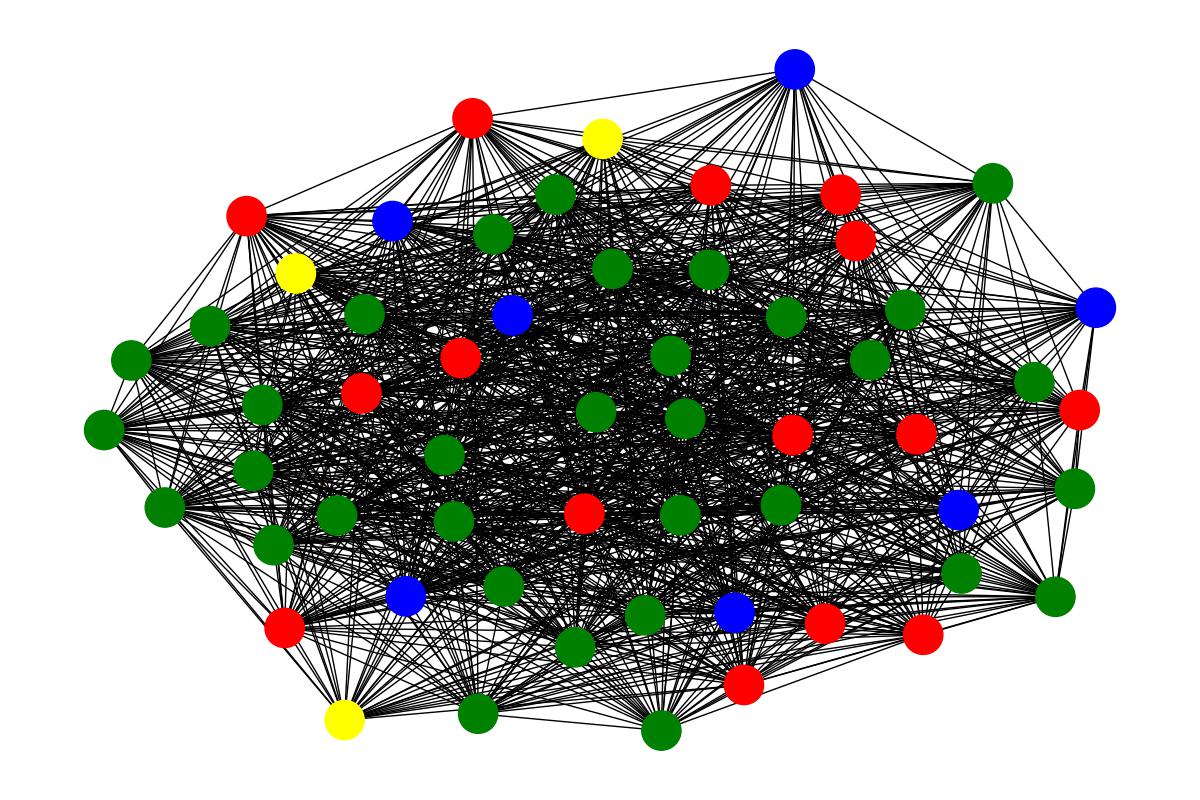}
        \label{fig:visual_fb_digress}
    }\hfill
    \subfigure[Generation of \algo.]{%
        \centering
        \includegraphics[width=0.3\textwidth]{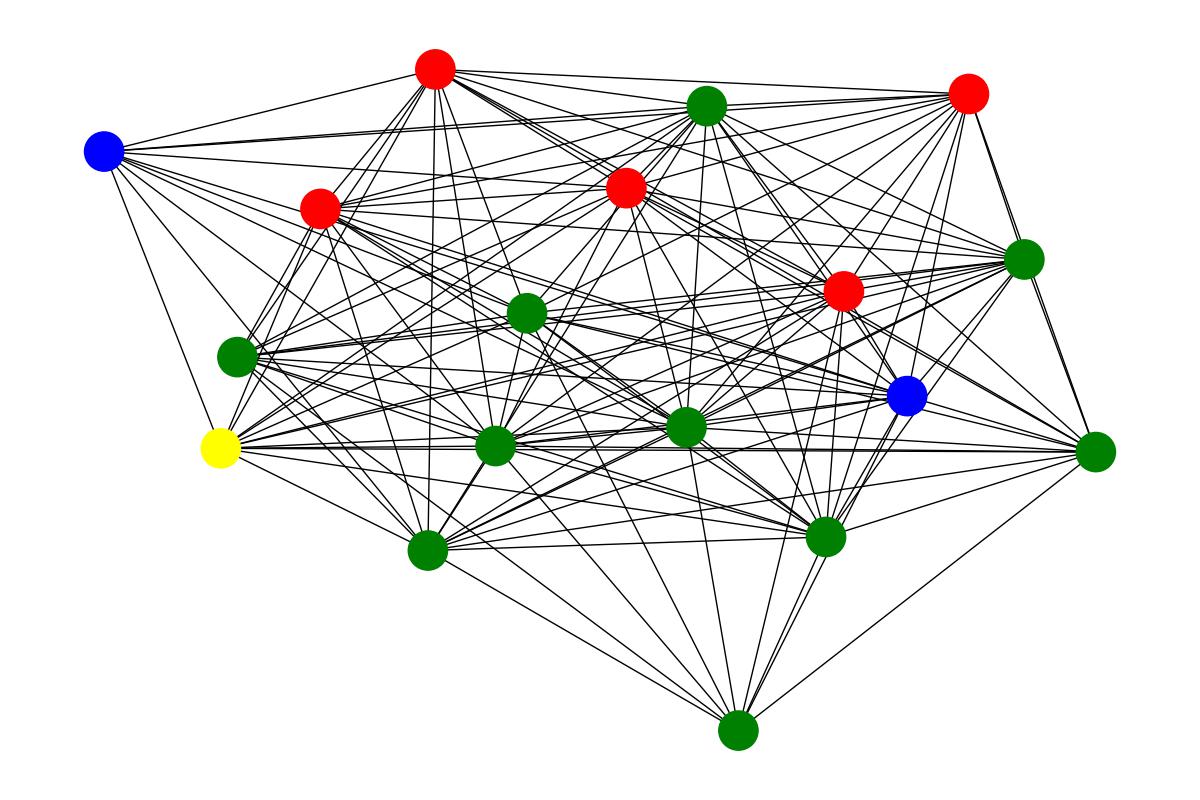}
        \label{fig:visual_fb_cdgraph}
    }\hfill%
     \caption{Visualization results of Facebook.}
     \label{fig:visual_fb}
\end{figure}

\end{document}